\documentclass[conference]{IEEEtran}

\ifCLASSINFOpdf

\else

\fi

\usepackage[cmex10]{amsmath}
\usepackage{tikz}
\usepackage{amsmath}
\usepackage{amssymb}
\usepackage{pifont}
\usepackage{times,url,color,soul,xspace,enumitem}
\usepackage{graphicx}
\usepackage{caption}
\usepackage{booktabs}
\usepackage{comment}
\usepackage{xspace}
\usepackage{tabularx}
\usepackage{pifont}       
\usepackage{bbding}      
\usepackage{fontawesome}  
\newcommand{\cmark}{\ding{51}}
\newcommand{\xmark}{\ding{55}}

\usepackage{tikz}
\usepackage[inline,draft,nomargin,index]{fixme}
\usepackage{multirow}
\usepackage{stmaryrd}
\usepackage{bm}
\usepackage{algorithmic}
\usepackage{array}
\usepackage{hyperref}
\usepackage{mdwmath}
\usepackage{mdwtab}
\usepackage[font=footnotesize]{subfig}
\usepackage{stfloats}
\usepackage{graphicx}
\usepackage{url}
\usepackage[misc]{ifsym}
\usepackage{latexsym}

\usepackage{cleveref}
\crefformat{section}{\S#2#1#3} % see manual of cleveref, section 8.2.1
\crefformat{subsection}{\S#2#1#3}
\crefformat{subsubsection}{\S#2#1#3}
\FXRegisterAuthor{giu}{agiu}{\textcolor{red}{Giu}}

\FXRegisterAuthor{atr}{aatr}{\textcolor{blue}{Atr}}

\FXRegisterAuthor{lin}{alin}{\textcolor{green}{Lin}}

\newcommand{\sys}{\textit{Fusion}\xspace}

\newtheorem{theorem}{Theorem}

\newtheorem{definition}{Definition}

\definecolor{brickred}{rgb}{0.8, 0.25, 0.33}
\usepackage{hyperref}
 
\hypersetup{
    colorlinks = true,
    linkcolor = red,
    anchorcolor = red,
    citecolor = red,
    filecolor = red,
    urlcolor = red
}
\newcounter{protocol}

\begin{document}

\title{\textit{Fusion}: Efficient and Secure Inference Resilient to Malicious Servers}

% \author{Caiqin Dong, Jian Weng, Jia-Nan Liu, Yue Zhang, Yao Tong, Anjia Yang, Yudan Cheng, and~ Shun Hu
% \thanks{C. Dong, J. Weng, J.-N. Liu, A. Yang, Y. Cheng, S. Hu are with the College of Cyber Security, the National Joint Engineering Research Center of Network Security Detection and Protection Technology, and also the Guangdong Key Laboratory of Data Security and Privacy Preserving, Jinan University, Guangzhou 510632, China. (Emails: caiqindong611@gmail.com, cryptjweng@gmail.com, j.n.liu@foxmail.com). 
% Yao Tong is with Guangzhou Fongwell Data Limited Company. Email: melody@fongwell.com.
% The corresponding author is Jian Weng.
%  % end author
% }
% }
% IEEEauthorblockN

\author{\IEEEauthorblockN{Caiqin Dong\textsuperscript{\dag}, Jian Weng\textsuperscript{\dag, \Letter}, Jia-Nan Liu\textsuperscript{\dag}, Yue Zhang\textsuperscript{\dag}, Yao Tong\textsuperscript{\S}, Anjia Yang\textsuperscript{\dag}, Yudan Cheng\textsuperscript{\dag}, and~ Shun Hu\textsuperscript{\dag}}
\IEEEauthorblockA{\textsuperscript{\dag}Jinan University, \textsuperscript{\S}Guangzhou Fongwell Data Limited Company\\
E-mails: \{caiqindong611, cryptjweng, zyueinfosec, anjiayang, gzhushun\}@gmail.com, \\\{j.n.liu, yudan\_cheng\}@foxmail.com, melody@fongwell.com}
}

\IEEEoverridecommandlockouts
\makeatletter\def\@IEEEpubidpullup{6.5\baselineskip}\makeatother
\IEEEpubid{\parbox{\columnwidth}{
  Network and Distributed System Security (NDSS) Symposium 2023\\
  27 February - 3 March 2023, San Diego, CA, USA\\
  ISBN 1-891562-83-5\\
    https://dx.doi.org/10.14722/ndss.2023.23199\\
    www.ndss-symposium.org
}
\hspace{\columnsep}\makebox[\columnwidth]{}}

\maketitle

\begin{abstract}
In secure machine learning inference, most of the schemes assume that the server is semi-honest (honestly following the protocol but attempting to infer additional information). However, the server may be malicious (e.g., using a low-quality model or deviating from the protocol) in the real world. Although a few studies have considered a malicious server that deviates from the protocol, they ignore the verification of model accuracy (where the malicious server uses a low-quality model) meanwhile preserving the privacy of both the server's model and the client's inputs. To address these issues, we propose \textit{Fusion}, where the client mixes the public samples (which have known query results) with their own samples to be queried as the inputs of multi-party computation to jointly perform the secure inference. Since a server that uses a low-quality model or deviates from the protocol can only produce results that can be easily identified by the client, \textit{Fusion} forces the server to behave honestly, thereby addressing all those aforementioned issues without leveraging expensive cryptographic techniques. Our evaluation indicates that \textit{Fusion} is 48.06$\times$ faster and uses 30.90$\times$ less communication than the existing maliciously secure inference protocol (which currently does not support the verification of the model accuracy). In addition, to show the scalability, we conduct ImageNet-scale inference on the practical ResNet50 model and it costs 8.678 minutes and 10.117 GiB of communication in a WAN setting, which is 1.18$\times$ faster and has 2.64$\times$ less communication than those of the semi-honest protocol.

\end{abstract}

% \begin{IEEEkeywords}
% Machine learning as a service (MLaaS), privacy-preserving inference, malicious server, verifiability.

% \end{IEEEkeywords}

% \IEEEpeerreviewmaketitle
% ******************************************
\section{Introduction}
\label{introduction}

Machine Learning as a Service (MLaaS) \cite{ribeiro2015mlaas,hesamifard2018privacy, CrankshawWZFGS17} is rapidly emerging as a dominant computing paradigm in the past few years.
In such a paradigm, the servers provide cloud-based machine learning services for the clients. As such, the clients now do not need to train their own models (which is computationally expensive, and requires large datasets), but consume the services on demand: the clients can simply feed the server inputs, and let the server make inferences based on the inputs. For example, a patient can provide the raw medical data to the server, and the server can then generate diagnosis results using the pre-trained models.

While MLaaS has brought the enormous convenience, it is also subject to privacy risks. For example, the inputs of clients may be highly sensitive (e.g., raw medical data) with confidentiality concerns~\cite{esteva2019guide}, and it is urgently necessary for the server to provide inference services without breaking their confidentiality.
Currently, many efforts have been made toward this end~\cite{demmler2015aby, patra2021aby2, jiang2018secure, liu2017oblivious, juvekar2018gazelle, chen2019efficient, riazi2019xonn, mishra2020delphi, rathee2020cryptflow2,rathee2021sirnn,0002SKG19, hazay2019leviosa, kumar2020cryptflow} via various cryptographic techniques such as homomorphic encryption (HE), garbled circuits (GC), and secret sharing (SS).
However, there could be plenty of large matrix multiplications, non-linear operations, and secure conversions back and forth between them in the neural network inference, which are relatively expensive to achieve with cryptographic techniques.
To achieve practical and privacy-preserving MLaaS, most of those works have no alternatives but to settle for less by adopting a weaker threat model --- assuming the servers and the clients will follow the protocol, but they will also try to obtain additional information (a.k.a, semi-honest security).

However, there is no reason to believe that such an assumption will always hold: the server can be completely malicious, and does not follow the protocols at all (e.g., the server returns the client random results to trick the clients)~\cite{ghodsi2017safetynets, zhao2021veriml,feng2021zen}.
Furthermore, even if the server follows the protocol, there is no guarantee that the server will produce high-quality results as promised, as few protocols can verify whether the inference results are produced by a high-quality model.
Those incorrect or inaccurate results can have grave consequences (e.g., misleading patients).
As such, solutions that solely consider the semi-honest security are obsolete, and we have to deal with a much worse scenario, where ($i$) the server is completely malicious (i.e., malicious security) to deviate from the protocol and  provide incorrect results, or ($ii$) although the server follows the protocol, it uses a low-quality model as input, and ($iii$) the server and client have the requirements of protecting their highly sensitive data (including the server's model and client's query input).
\looseness=-1

To our best knowledge, there is no solution that can satisfy all the criteria mentioned above. For example, 
\textit{LevioSA} \cite{hazay2019leviosa} achieves the maliciously secure arithmetic computation by following the high-level approach of the IPS compiler~\cite{ishai2009secure,lindell2011ips} (IPS compiler is designed to construct secure protocols in the presence of malicious adversaries) and applies it to privacy-preserving machine learning.  
C\textsc{ryp}TF\textsc{low}~\cite{kumar2020cryptflow}  converts \textsf{TensorFlow}~\cite{abadi2016tensorflow} inference code to MPC protocols.
By using the trusted hardware, their solution also satisfies malicious security.
Although those solutions~\cite{kumar2020cryptflow, hazay2019leviosa} guarantee computation correctness (which ensures the server follows the protocol correctly) and preserve privacy, they do not verify the model accuracy (e.g., the malicious server could still use a low-quality model). Moreover, some schemes \cite{ghodsi2017safetynets, lee2020vcnn,zhao2021veriml,feng2021zen,liuzkcnn,weng2021mystique} utilize zero-knowledge (ZK) proofs to compel the server to provide correct inference results for the client, but they only protect the privacy of either the server's model or the client's input.

This paper aims to propose a solution for MLaaS that satisfies all three criteria (including verification of model accuracy, computation correctness, and privacy preservation) at the same time. To that end, there could be multiple ways. For example, we can directly apply maliciously secure two-party computation (2PC), which preserves privacy and ensures computation correctness, but extra expensive cryptographic approaches (e.g., ZK proofs) are required to verify the model accuracy. Alternatively, 
we can make changes on the existing efficient semi-honest inference schemes~\cite{liu2017oblivious, juvekar2018gazelle, mishra2020delphi} to achieve malicious security. Given that these schemes usually use multiple cryptographic techniques such as HE and GC simultaneously, it is also challenging to customize them to let them meet all three criteria (particularly, enabling the verification of model accuracy may require significant modifications) effectively.\looseness=-1

Fortunately, we observe that the client can know the computation results (public samples' labels) of some inputs (or query samples) in advance, which is not possible in most secure computation applications. These pre-collected public samples with known (or expected) computation results can be used to verify the model accuracy meanwhile force the server to perform  computations correctly by mixing them with real query samples.
Based on this observation, we customize a \textit{mix-and-check} method that combines the verification of model accuracy with computation correctness for batched inference queries. Specifically, we design a mixed dataset by preparing query samples (non-public, and each with multiple copies) and a number of public samples, and then using a random permutation to shuffle them. 
If the server attempts to cheat the client without being noticed, the server has to provide incorrect-but-consistent results for all copies of a particular query sample. Given that the server will not know how the samples (the public ones and non-public ones) are mixed according to our design, it is extremely challenging for the server to cheat successfully, and the clients can easily notice the misbehavior with overwhelming possibility.

By using this \textit{mix-and-check} method, we propose a maliciously secure inference scheme, named \sys, which can convert a semi-honest inference protocol into a maliciously secure one.
\sys is superior because it fulfills the aforementioned three security requirements effectively.
It preserves the privacy of both the server and the client by utilizing a semi-honest secure inference protocol, and ensures the computation correctness and model accuracy effectively (it uses simple-but-effective local checks, not expensive cryptographic techniques).
We have also implemented \sys and compared its performances with existing maliciously secure solutions and semi-honest inference protocols.
Moreover, we conduct \textsf{ImageNet-scale} inference on practical \textsf{ResNet50} model.
When compared with C\textsc{ryp}TF\textsc{low}2 \cite{rathee2020cryptflow2},
\sys is 1.30$\times$ and 1.18$\times$ faster in the LAN setting and the WAN setting respectively, and uses 2.64$\times$ less communication (when the total number of query samples is 512 and the copies for each query sample is 5 that ensure the statistical security of $2^{-40}$).  
Finally, we also enable \sys to defend against model extraction attacks (by integrating a prior solution), which proves that \sys has good scalability.   

 In short,  our contributions are twofold: 
 \begin{itemize}
     \item  First, we propose \sys, a maliciously secure inference scheme. To our best knowledge, \sys is the first solution that preserves the privacy of both the server and the client, and ensures the model accuracy of the server's input and computation correctness at the same time in the \textit{server-malicious} threat model.
    Particularly, \sys can be used as a general compiler that converts a semi-honest inference scheme into a maliciously secure one.
    As a consequence, the proposed scheme can benefit from the existing efficient and practical inference schemes. 
    \item  Second, we implement \sys, and compare its performance with the state-of-the-art maliciously secure work \textit{LevioSA}~\cite{hazay2019leviosa}.
    The results are encouraging: \sys is 48.06$\times$ faster and uses 30.90$\times$ less communication than \textit{LevioSA}~\cite{hazay2019leviosa} (which currently does not support verification of the server's model accuracy).
    We also compare \sys with multiple semi-honest inference protocols such as DELPHI~\cite{mishra2020delphi} and ABY~\cite{demmler2015aby}, and better performance is also observed on \sys when they all adopted the same settings (e.g., the number of query samples is large enough such as 32). 

 \end{itemize}

%******************************************
\section{Preliminaries and Background}
\label{preliminaries}
In this section, we describe some background information about neural network inference, followed by privacy-preserving neural network inference, where we introduce a few popular  hybrid 2PC-based privacy-preserving inference protocols. 
\looseness=-1

\subsection{Neural Network Inference}\label{NNI}
Being one of the important types of deep learning, convolutional neural network (CNN) inference computations mainly contain four components, i.e., convolutions, which extract different features from a dataset, pooling layer, which mainly attempts to reduce the dimensionality, activation function, which is used as a feature transformation method to increase the expression ability, and fully connected layer, which takes the outputs of other components and produces the final outputs. All those components are connected, and ultimately form a multi-layer network structure, where the outputs of one layer can be the inputs of other layers. CNN inference usually contains a lot of computationally expensive linear and  non-linear computations. For example, convolutions usually contain multiple {linear} matrix multiplications; activation functions such as rectified linear unit (ReLU)~\cite{nair2010rectified} are nonlinear transformations. Pooling functions (e.g., max pooling functions) usually are nonlinear operations as well.

\subsection{Privacy-Preserving Neural Network Inference}
\label{ppnni}

In MLaaS without considering privacy preservation, the client's inputs are directly sent to the server who produces the inference results, and the highly sensitive inputs (e.g., raw medical test results) can be leaked to the server. As such, the privacy-preserving neural network inference is introduced,  which generates the outputs without leaking sensitive inputs to the server.
At a high level, the client and the server adopt cryptographic techniques to perform inference computations for protecting their own inputs.
Among all the privacy-preserving inference frameworks, semi-honest privacy-preserving inference gained popularity due to its practical efficiency. 

An intuitive way to implement privacy-preserving inference is through two-party computation (2PC). 2PC is a sub-problem of secure multi-party computation (MPC), which allows two parties to jointly compute a function without sacrificing their input privacy. However, due to the complexity of operation types (e.g., nonlinear comparison operations, matrix multiplication) and plenty of conversions between these operations in neural networks, directly using general 2PC schemes faces efficiency challenges.

Having seen the efficiency challenges, researchers proposed hybrid 2PC-based privacy-preserving inference protocols~\cite{liu2017oblivious, juvekar2018gazelle, mishra2020delphi, HuangLHD22}, which improve the overall efficiency by using appropriate cryptographic techniques for different kinds of computation or designing new efficient subprotocols.
In the following, we would like to introduce a few popular hybrid 2PC-based privacy-preserving inference frameworks:

\begin{itemize}
  \vspace{1mm}
    \item \textbf{ABY}~\cite{demmler2015aby} is a semi-honest mixed-protocol framework that combines Arithmetic (A) sharing, Boolean (B) sharing, and Yao’s (Y) garbled circuits, and designs efficient conversions between every pair of the three.
    It also adopts a set of existing state-of-the-art optimizations in a novel fashion and provides instantiations.
    
    \vspace{1mm}
    \item  \textbf{DELPHI}~\cite{mishra2020delphi}
     combines the additive homomorphic encryption with garbled circuits (HE is used to perform linear matrix multiplication while GC is used to perform non-linear operations) and connects them by additive secret sharing (e.g., the output of the linear layer are additive secret shares of computation results that are ultimately fed into the GC for non-linear operations).
     Besides, it allows users to automatically generate neural networks that mix these two methods and navigate the trade-off between accuracy and performance.

     \vspace{1mm}
    \item  \textbf{C\textsc{ryp}TF\textsc{low}2}~\cite{rathee2020cryptflow2} is also a hybrid inference protocol that shares some similarity with DELPHI.
    To improve efficiency, it proposes a more efficient millionaires protocol $\Pi_{\text{MILL}}^{l,m}$ for securely computing the Yao's millionaires' problem (which is used for non-linear operations such as DReLU and Maxpool). C\textsc{ryp}TF\textsc{low}2 also provides two options, i.e., $\text{SCI}_{\text{OT}}$ and $\text{SCI}_{\text{HE}}$ that use oblivious transfer and homomorphic encryption techniques to perform computationally costly linear operations such as matrix multiplication or convolution. 
    
    \vspace{1mm}
    \item \textbf{Cheetah}~\cite{HuangLHD22} is also a hybrid secure inference scheme that utilizes two lattice-based homomorphic encryptions~\cite{chen2021efficient} (i.e., learning with errors (LWE) and its ring variant (ring-LWE)) to perform secure linear layers (e.g., matrix multiplications in convolution), and makes some optimizations on the millionaire protocol \cite{rathee2020cryptflow2} for non-linear layer (e.g., activation function).
    They achieve performance optimizations based on the insightful observation that matrix multiplication results can be represented as the coefficients in specific positions of polynomial multiplication, which can be efficiently performed using ring-LWE.
    For the non-linear layers, they optimize the millionaire protocol by adopting VOLE-style OT and customizing truncation protocols.
 
\end{itemize}

%-------------------------------------------------------------------------------
%******************************************
\section{Threat Model and Security Goals}
\label{systemmodel}
\begin{figure}[!ht]
  \centering
  \includegraphics[width=0.8\linewidth]{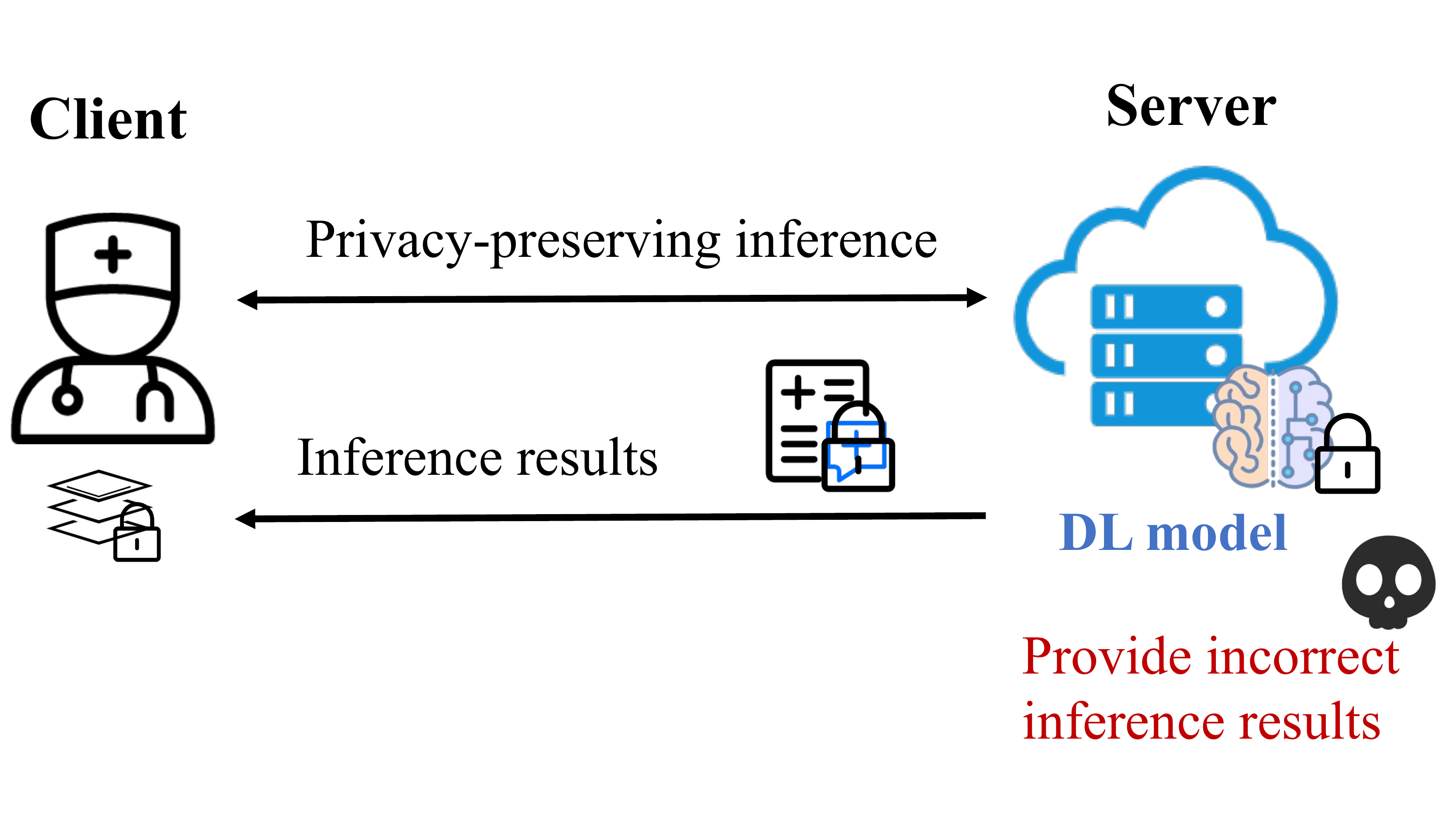}\\
  \caption{System model.}
  \label{model}
\end{figure}

In privacy-preserving neural network inference, the server and the client jointly perform secure 2PC inference computations with their private inputs (e.g., model parameters or query samples). As such, similar to all other privacy-preserving neural network inference, we consider two entities in the system model, as shown in \autoref{model}: The server owns a well-trained (in terms of accuracy) model for a specific task (e.g., medical diagnosis) and provides inference services to the clients. The client owns a set of query samples, and it attempts to obtain the correctly computed inference results from the server. During the process, the server is required to use a well-trained model as input, and the client uses query samples as inputs for privacy-preserving inference computations. 
In the rest of this section, we describe the threat model and security requirements.

\subsection{System Model and Threat Model}
\label{subsec:threatmodel}

Similar to previous efforts (e.g.,  \cite{feng2021zen,weng2021mystique, LehmkuhlMSP21}), in this paper, we consider the server is malicious that may not follow the protocol while the client is honest. In particular, given there are two roles (i.e., the server and the client), we make the following assumptions for each of them respectively:

\vspace{1mm}
\begin{itemize}
\item \textbf{Server:} We assume that the server may trick the clients by using a low-quality model (in terms of accuracy) as input. Meanwhile, the server may deviate from the protocol specification. However, the server still aims to convince the client that the inference results are correctly computed from the client's query samples using a high-quality model (e.g., high accuracy). 

\vspace{1mm}
\item \textbf{Client:}
The client honestly follows the protocol but may attempt to obtain additional information (i.e., semi-honest). The client also holds a set of samples to be queried. We additionally assume that the client has access to a subset of public samples whose type is the same as its query samples. This is reasonable since there are plenty of public datasets with different categories (e.g., images, audio, and healthcare data) that can be used for various purposes (e.g., face recognition and medical diagnosis).
In particular, we can even find publicly available datasets that are as sensitive as medical records on some large national databases (e.g., TGGA~\cite{TCGA} and NCBI~\cite{NCBI}) or institutes (e.g., AIMI~\cite{AIMI}), as data sharing can help accelerate disease research and improve diagnostic methods. Examples of those medical records datasets include genomic data~\cite{NCBIgene, TCGAbrca}, gene expression profiles~\cite{gene, NCBIgeo}, COVID-19-Data~\cite{COVID, wang-etal-2020-cord}, and medical image data~\cite{AIMI, tmi.2020.2992244-19}. As such, it is practical for clients to obtain a number of public samples for use.

\end{itemize}

\subsection{Security Goals}
\label{subsec:securitygoals}

While the semi-honest assumptions (by assuming the server and the client will honestly follow the protocol without cheating and other misbehavior) are practical and widely adopted by a lot of previous efforts, there is no reason to believe that such an assumption will result in expected outputs. For example, the server may be malicious, which feeds the protocol low-quality model to trick the clients as discussed in \S\ref{subsec:threatmodel}. As such, this paper considers a much stronger threat model, which attempts to meet at least the following security goals: (1) we should not leak the input privacy of both the server and the client; (2) we should ensure the computation correctness of the outputs; and (3) we should ensure the accuracy of the inference results (e.g., the server uses high-quality models as inputs).

\noindent\textbf{Formalization. } Without loss of generality, we provide security in the simulation paradigm.
We consider a \textit{hybrid model} where parties both interact with each other and have access to ideal functionalities.
Assume that a two-party hybrid-model protocol $\pi$ uses ideal calls to ideal functionalities $f_1, \cdots, f_{p(n)}$.
Let $\rho_1, \cdots, \rho_{p(n)}$ be protocols that securely compute $f_1, \cdots, f_{p(n)}$ respectively.
The composition theorem~\cite{Canetti00} states that if $\pi$ securely computes the functionality $g$ in the $(f_1, \cdots, f_{p(n)})-$hybrid model, then $\pi^{\rho_1, \cdots, \rho_{p(n)}}$ in which the ideal functionalities are substituted by the secure sub-protocols securely computes $g$ in the real model.

\begin{definition}\label{securitydefinition}
 \textit{A protocol $\Pi_{\sys}$ between a server having model $\mathrm{M}$ which satisfies the accuracy threshold $\delta$ and a client having a dataset $\mathrm{X}=(\mathrm{x}_1,\cdots, \mathrm{x}_n)$ as inputs securely achieves a secure inference functionality $\mathcal{F}_{\sys}$ against a malicious server and a semi-honest client if it satisfies the following requirements:}
\end{definition}

  \begin{itemize}
    \item \textbf{\textit{R1.~Model Accuracy}.}
    \textit{The accuracy (e.g., $\eta$) of the model that the server uses as input should meet the requirement (e.g., $\delta$).
    Assume that the model accuracy $\eta$ is calculated. If $\eta \geq \delta$, the verification of model accuracy passes.}\\

    \item \textbf{\textit{R2.~Computation 
    Correctness}.}
    \textit{In an execution of $\Pi_{\sys}$, the probability that the client's output on every input vector $\mathrm{x}_i$ is not the correct inference result $\mathrm{M}(\mathrm{x}_i)$ is negligible in security parameter $\lambda$.}\\

    \item \textbf{\textit{R3.~Privacy}.} \textit{The server and the client both feed their private inputs to $\Pi_{\sys}$. As such, from the perspective of privacy preservation, the client and the server is said to securely execute  $\Pi_{\sys}$ against a malicious server and semi-honest client if the following properties are satisfied:}
    
    \begin{itemize}
        \item \textbf{\textit{Malicious Server Security}.}
    \textit{The view of the server during a real execution of protocol $\Pi_{\sys}$ is denoted by $\textsf{View}_{\mathcal{S}}^{\Pi_{\sys}}$.
    For any server $\mathcal{S}$, there exists a probabilistic polynomial-time simulator $\textsf{Sim}_{\mathcal{S}}$ such that for any input $\mathrm{M}$ of the server and $\mathrm{x}_i$ of the client, we have:\\
    \centerline{$\textsf{View}_{\mathcal{S}}^{\Pi_{\sys}}\approx_{c} \textsf{Sim}_{\mathcal{S}}(\mathrm{M})$}}
    
{\textit{That is, $\textsf{Sim}_{\mathcal{S}}$ can simulate a computationally indistinguishable view of the malicious server without knowing the client's private inputs and inference results.}} 

\item \textbf{\textit{Semi-honest Client Security}.}
\textit{The view of the client during a real execution of protocol $\Pi_{\sys}$ is denoted by $\textsf{View}_{\mathcal{C}}^{\Pi_{\sys}}$.
For any client $\mathcal{C}$, there exists a probabilistic polynomial-time simulator $\textsf{Sim}_{\mathcal{C}}$ such that for any input $\mathrm{M}$ of the server and $\mathrm{x}_i$ of the client, we have:\\
\centerline{$\textsf{View}_{\mathcal{C}}^{\Pi_{\sys}}\approx_{c} \textsf{Sim}_{\mathcal{C}}(\mathrm{x}_i,\mathrm{M}(\mathrm{x}_i))$}
That is, $\textsf{Sim}_{\mathcal{C}}$ can simulate a computationally indistinguishable view of the semi-honest client without knowing the server's model.}
\end{itemize}
\end{itemize}

\section{The \textit{Fusion} Scheme}
\label{scheme}

\begin{figure*}[!ht]
  \centering
  \includegraphics[width=0.85\linewidth]{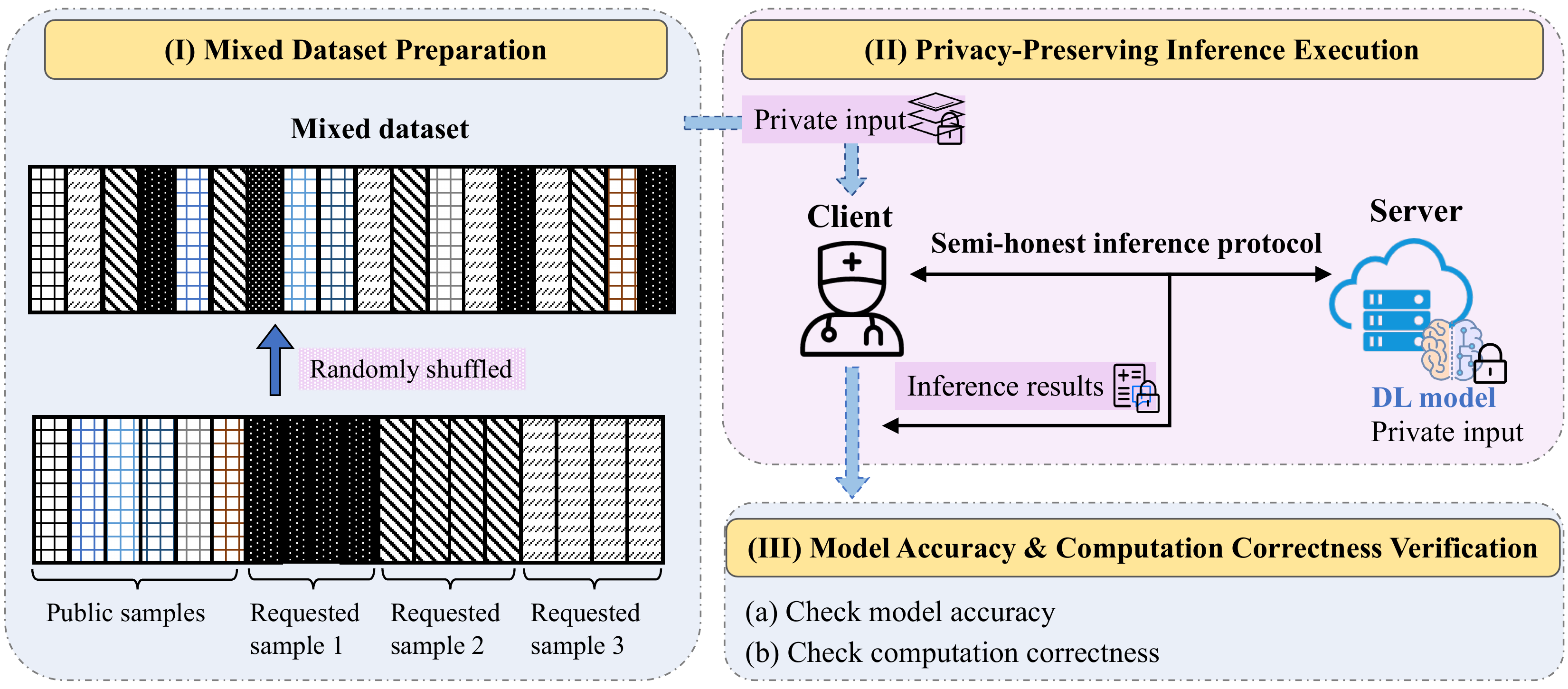}\\
  \caption{The workflow of \sys. Particularly, for better illustration, we highlight the local operations of the client as blue, and highlight the interaction between the server and the client as pink.}
  %The client first locally prepares a mixed dataset. Next, the server and the client jointly perform semi-honest privacy-preserving inference computations without revealing respective inputs. After completing the inference, the results are only revealed to the client, with which it locally checks the server's model accuracy and computation correctness .}
  \label{SystemModel}
  %\vspace{-0.2in}
\end{figure*}

In this section, we propose \sys, a privacy-preserving inference scheme that is secure against a malicious server and a semi-honest client. We first explain our intuition, then we present the overview of \sys, and finally, we describe the current design in greater details. 

\subsection{Challenges and Key Ideas}
\label{keyideas}

As discussed in \S\ref{subsec:securitygoals}, there are three goals that need to be fulfilled in the presence of a malicious server, i.e., model accuracy, computation correctness, and privacy.
To our best knowledge, there is no existing work that satisfies all three goals mentioned above in neural network inference.
The latter two requirements (R2 and R3) can be achieved by directly adopting general maliciously secure protocols (e.g., popular SPDZ-style protocols~\cite{DamgardPSZ12, keller2018overdrive}), though they are not efficient enough for performing neural network inference.
In addition, maliciously secure protocols do not provide guarantees for the model accuracy (e.g., those protocols cannot force the server to use a high-quality model as input). Even though there are some additional techniques that could potentially be used to verify model accuracy, they usually involved heavy cryptographic techniques (e.g., using ZK proofs and commitments~\cite{weng2021mystique} to convert publicly committed data into privately authenticated values).
As such, it is challenging to achieve all three goals in neural network inference efficiently.

Our goal is to achieve all three security requirements without using heavy cryptographic techniques such as ZK proofs.
% (e.g., we will not consider SPDZ-style protocols, as they are too heavy for neural network inference).
Generally, the privacy requirement (R3) can be achieved by adopting efficient 2PC protocols to perform the neural network inference. We now discuss how to fulfill model accuracy (R1) and model correctness (R2) at the same time.  Our idea is that if we can know the outputs of specific inputs in advance,  we can use such knowledge to ensure the \textit{computation correctness} and \textit{model accuracy}. This is because if the server causes wrong results (violating R2) or uses low-quality inputs (violating R1), the client can notice that by comparing the expected outputs (which are known in advance) with the actual outputs (which are obtained at real time), thereby forcing the server to behave honestly. 
Particularly, knowing the outputs in advance is possible, since there are samples (e.g., TGGA~\cite{TCGA} and NCBI~\cite{NCBI} databases) where both the inputs and outputs (labels) are publicly available, as discussed in \S\ref{subsec:threatmodel}.  
The key question now becomes how we can find a way to appropriately utilize these public samples, and feed them into the server to achieve R1 and R2.

The first solution that comes to our mind is 
\textit{cut-and-choose} technique~\cite{lindell2007efficient},  a leading technique used to convert Yao's garbled circuit \cite{yao1986generate} to be maliciously secure: in our scenarios, we need a method to pinpoint the server's malicious behavior, while the insight behind the \textit{cut-and-choose} technique is that if a malicious generator (which takes responsible for constructing garbled circuits for computing a function, such as a server) constructs incorrect garbled circuits, the evaluator (which evaluates the garbled circuits to obtain outputs, such as a client) would detect the malicious behavior with high probability.
We now provide more details of batched \textit{cut-and-choose} (that can amortize the cost across many executions of the same protocol): the generator first constructs many garbled circuits and sends them to the evaluator, then the evaluator asks the generator to open some of the garbled circuits (e.g., revealing the decryption keys corresponding the chosen garbled circuits) for checking the correctness of these garbled circuits (meaning that the circuit indeed computes the expected function).
If all opened circuits are correctly constructed, the remaining circuits are randomly grouped into various buckets of a specific size, and all garbled circuits in the same bucket are evaluated with the same inputs from two parties.
As a result, the evaluator would identify the malicious behavior (by checking the construction correctness of opened circuits and the result consistency of all copies in the same bucket) with high probability if the generator constructs incorrect garbled circuits.

However, the \textit{cut-and-choose} strategy cannot be directly applied to compile the hybrid 2PC-based inference protocol (which typically utilizes additive secret sharing to connect two different cryptographic protocols such as HE and GC) to a maliciously secure one. The reasons for that are multiple: First, the \textit{cut-and-choose} technique for GC only ensures the correctness (meaning that the garbled circuit is correctly constructed for computing the expected function). However, in our case, we need to verify the server's inputs (not the produced garbled circuits) which the \textit{cut-and-choose} technique itself cannot achieve.
Second, the \textit{cut-and-choose} technique directly opens the generated circuits for checking correctness because the garbled circuits themselves do not relate to specific inputs (if they are never used for evaluation), and therefore can be opened publicly without sacrificing the privacy of both parties' inputs.
However, in our scenarios, we customize the \textit{mix-and-check} method to convert semi-honest inference protocols into maliciously secure counterparts, while the underlying hybrid 2PC-based inference protocols generally do not have the structure that can be directly opened for checking correctness without sacrificing input privacy.\looseness=-1

\vspace{1mm}
\noindent\textbf{Key Insight.} Fortunately, inspired by the \textit{cut-and-choose} technique, we propose a novel technique named \textit{mix-and-check}, which forms our technical foundation, and allows us to overcome all challenges above. Our idea is to let the client mix a set of public samples (imitating the evaluator's choice of randomly opening some garbled circuits) together with the samples to be queried, and use them as inputs for executing  privacy-preserving inference (R3). 
If the client observes any inconsistency (e.g., the outputs of these public data records do not equal to their expected outputs, or inconsistent results are observed across some of the copies), we detect that the server does not honestly use a high-quality model or the server does not follow the specified protocol.
These public samples are used to verify the model accuracy (R1) and computation correctness (R2) at the same time in a novel fashion.

\subsection{Overview of \sys}

Based on our newly proposed \textit{mix-and-check} method, we design a new maliciously secure inference protocol, \sys.  We assume that before the server and client run \sys, the server trains a model and uses it to provide the inference services. As shown in \autoref{SystemModel}, \sys works as follows: 

\begin{enumerate}
  
\item \textbf{Mixed Dataset Preparation.} 
The client locally prepares a mixed dataset (with public samples, and the samples to be queried) and utilizes the mixed dataset as input for the privacy-preserving inference computations.
\item \textbf{Privacy-Preserving Inference Execution.}  The server and the client jointly perform the semi-honest privacy-preserving inference protocols, and the inference results are only revealed to the clients.

\item \textbf{Model Accuracy and Computation Correctness Verification.} The client verifies the model accuracy on public samples,  and computation correctness through the consistency check of inference results for all copies of every query sample.
The client accepts the inference results if the above two checks pass. 

\end{enumerate}

\subsection{Detailed Design of \sys}
\label{subsec:designsys}

We summarize the maliciously secure inference protocol $\Pi_{\textit{Fusion}}$ in \autoref{Fusion} and describe the process in detail as follows.

\noindent\textbf{(I) Mixed Dataset Preparation.} 
In this phase, the client locally prepares the  {\textit{mixed dataset}}.
To that end, the client first selects $R$ \textit{query samples}, duplicates $B$ copies of every \textit{query samples} (in total, there are $R * B$ copies), and prepares $T$ \textit{public samples} for future inference computations. Next, the client uses a randomly chosen permutation to shuffle all public samples and copies of all query samples.

The random shuffle of the query samples and public samples is easy to design and implement, and therefore, in the following, we would like to discuss how to select optimal $B$ and $T$. Particularly, we have two requirements. 
First, the server should not successfully trick the client into believing that the incorrect or low-quality results are correctly produced by a high-quality model. 
The statistical security parameter is denoted by $\lambda$. 
The security requirement is to guarantee that the server succeeds in cheating with a probability (which depends on specific $R$, $B$, and $T$) at most $2^{-\lambda}$. We denote this requirement as \textit{security requirement}.
Second, in order to achieve the best efficiency, the goal of this phase is to pick the concrete numbers of $B$ and $T$ (given a specific $R$) that minimize the cost per query sample while satisfying the \textit{security requirement}.
We denote this requirement as the \textit{cost requirement.}
Specifically, computation and communication costs per query sample is proportional to the number of public samples $T$ and the number of copies $B$ for each query sample.
As such, we have to solve a parameter optimization problem that satisfies the \textit{security requirement} meanwhile meeting the \textit{cost requirement}.
\looseness=-1

We are now starting with the \textit{security requirement}.
The server can succeed in cheating the client when (1) the server passes the \textbf{\textit{accuracy check of public samples}}: accuracy is calculated based on the $T$ public samples; and (2) the server also passes the \textbf{\textit{consistency check of query samples}}: the inference results of all copies for every query sample are consistent and the inference results of some query samples are incorrect (produced by a low-quality model or incorrect inference computations). Without loss of generality, we further define those two types of passes mathematically: 
\begin{itemize}
    \item \textit{\textbf{Accuracy Check of Public Samples}:} To simplify the problem, we assume that the server knows $T$ and $B$.
Assume that the server attempts to corrupt $i$ query samples by providing $iB$ incorrect-but-consistent inference results for all copies of every query sample.
If the server attempts to pass the check of model accuracy (accuracy check of public samples), it should use the high-quality model as input for the $T$ public samples.
Let $E_T$ denote the event in which the server uses a high-quality model to correctly perform the inference computations on all $T$ public samples. As such, we have 
the probability Pr[$E_T$] that event $E_T$ happens: 

\begin{equation}
  \label{p2}
  \begin{aligned}
  \text{Pr}[E_T]&= \frac{\dbinom{RB+T-iB} {T} }{\dbinom{R B+T} {T}} \\&=\frac{(R B+T-iB)!(RB)!}{(RB-iB)!(RB+T)!}.
  \end{aligned}
\end{equation}

\item \textit{\textbf{Consistency Check of Query Samples}:} Let $E_B$ denote the event in which the $iB$ incorrect inference results chosen by the server are exactly the incorrect-but-consistent results for $i$ query samples.
There are $(RB)!$ ways to permute the $RB$ query samples.
Again, if the server attempts to cheat the client successfully, it should provide incorrect-but-consistent inference results for $iB$ copies of the $i$ query samples, and use the high-quality model to perform correct inference computations for the remaining samples.
Similarly, the probability $\text{Pr}[E_B]$ that event $E_B$ happens is as follows:

\begin{equation}
  \label{p3}
  \text{Pr}[E_B]= \frac{\dbinom{R}{i}( i B )!(RB-iB)!}{(RB)!}.
\end{equation}
\end{itemize}

\begin{figure*}[!ht]
  \fbox{\parbox[c][][t]{17cm}{
 \vspace{2mm}
\textbf{Input:} The server inputs a model $\mathrm{M}$ for a specific inference task, the client inputs $R$ query samples $\mathrm{X}=(\mathrm{x}_1, \cdots, \mathrm{x}_{R})$ and $T$ public samples $\{ (\mathrm{x}_1^{*}, y_1^{*}), \cdots, (\mathrm{x}_T^{*}, y_T^{*})\}$, and an accuracy threshold $\delta$ is set.\\
\textbf{Output:} The inference results $\mathrm{M}(\mathrm{x}_i^{\prime})$ ($i \in \{1, \cdots, RB+T\}$).

\begin{enumerate}

  \item \textbf{Mixed Dataset Preparation (only the client).}
  
  \begin{enumerate}
  \item Searches optimal $(B, T)$ that minimize $\textit{cost}(B,T,R) = \frac{RB+T}{R}$ by using the search protocol $\Pi_{\textit{Search}}$ (\autoref{optimizedparameters}).
  \item Duplicates $B$ copies for each query sample to obtain $\{(\mathrm{x}_1^{1}, \cdots, \mathrm{x}_1^{B}), \cdots, (\mathrm{x}_R^{1}, \cdots, \mathrm{x}_R^{B})\}$, uses a random permutation $\pi$ to mix them together with $(\mathrm{x}_1^{*}, \cdots, \mathrm{x}_{T}^{*})$, and finally obtains a mixed dataset $\mathrm{X}^{\prime}=(\mathrm{x}_1^{\prime}, \cdots, \mathrm{x}_{RB+T}^{\prime})$.
  \end{enumerate}

  \item \textbf{Privacy-Preserving Inference Execution.}
  
  Using $\{\mathrm{x}_1^{\prime}, \cdots, \mathrm{x}_{RB+T}^{\prime}\}$ and $\mathrm{M}$ as inputs respectively, the client and the server jointly invoke the semi-honest inference protocol and reveal the computation results $\mathrm{M}(\mathrm{x}_i^{\prime}) (i \in \{1, \cdots, RB+T\})$ to the client.

  \item \textbf{Model Accuracy and Computation Correctness Verification (only the client).}

  \begin{enumerate}
    \item For $i \in \{1, \cdots, T\}$, if $\mathrm{M}(\mathrm{x}_i^{*}) = y_i^{*} $, sets $y_i = 1$.
    Calculates the \textbf{\textit{model accuracy}} $\eta$ as follows.

    \begin{equation}
      \eta=\frac{\sum{y_i}}{T}.\nonumber
    \end{equation}
    
    If $\eta \textless \delta$, the client aborts.

    \item 
    Verifies the \textbf{\textit{computation correctness}} by checking whether $\{ \mathrm{M}(\mathrm{x}_i^{1}), \cdots, \mathrm{M}(\mathrm{x}_i^{B}) \} (i \in \{1, \cdots, R\})$ are all the same.
    If there is any inconsistency, the client aborts.
    
  \end{enumerate}

\end{enumerate}
            }
           }
            \caption{\textbf{Protocol $\Pi_{\textit{Fusion}}$ for batched secure inference}.}
            \label{Fusion}
            %\vspace{-0.2in}
\end{figure*}

Given that the server can succeed in cheating when it can pass both \textit{accuracy check of public samples} and \textit{consistency check of query samples} at the same time, combining \autoref{p2} and \autoref{p3}, the probability $\text{Pr}_{success}$ that the server succeeds in cheating is as follows: 
\begin{equation}
  \begin{aligned}
  \label{p4}
  \text{Pr}_{success}&= \text{Pr}[E_T \wedge E_B] \\&=\text{Pr}[E_T] \times \text{Pr}[E_B] \\ 
  &=\dbinom{R} {i}\dbinom{RB+T}{iB}^{-1}. 
  \end{aligned}
\end{equation}

As long as $ T \geq B$, we have the following equation (which is proved in \S\ref{securityanalysis}):   
\begin{align*}
\dbinom{R} {i}\dbinom{RB+T}{iB}^{-1} \leq R\dbinom{RB+T}{B}^{-1}
\end{align*}

The \textit{security requirement} states that $\text{Pr}_{success}$ should be no more than $2^{-\lambda}$ for every choice of $i$ by the server and of $R$, $B$, and $T$ by the client.

On the basis of satisfying the \textit{security requirement}, another goal is to find the optimal $B$ and $T$ that meet the \textit{cost requirement} (e.g., leading to the lowest amortized cost per query sample).
We denote a cost function $\textit{cost}(B,T,R) = \frac{RB+T}{R}$ which denotes the amortized cost per query sample.
Specifically, the parameter optimization problem must ensure the security constraint $\text{Pr}_{success} \leq 2^{-\lambda}$, and should try to minimize the cost function.
We additionally set a lower bound $\beta$ of the number of public samples for ensuring the reliability of the accuracy check of public samples.
Therefore, the parameter optimization problem can be expressed as follows.
\begin{equation}
  \label{p6}
  \mathop{\arg\min}_{B,T}\frac{RB+T}{R},
\end{equation}
subject to
\begin{equation}
  \label{p7}
   T\geq \beta,
\end{equation}

\begin{equation}
  \label{p8}
  \text{Pr}_{success}\leq 2^{-\lambda}.
\end{equation}

To find the optimal $B$ and $R$ that satisfy the \textit{security requirement} with minimized amortized cost, we design a search algorithm (shown in \autoref{optimizedparameters}) based on the probability constraint and cost function. 
For a given $R$, we search for a pair $(B, T)$ that minimizes the cost function $\textit{cost}(B, T, R)=\frac{RB+T}{R}$ while satisfying the \textit{security requirement} (e.g., $\text{Pr}_{success}\leq 2^{-\lambda}$).
Specifically, for every choice of $B$ ranging from 2 to $\lambda$, we search for $T$ for a given $B$ until we find the smallest $T$ that satisfies the \textit{security requirement} (the amortized cost decreases as $T$ decreases).
%To identify and update the optimal pair $(B^{*}, T^{*})$, 
We continue to explore and update the optimal pair $(B^{*}, T^{*})$ with the current pair $(B, T)$ if the current pair saves more of the amortized cost (according to the cost function) than the optimal pair.
Finally, we obtain the optimal $(B, T)$ that minimizes the amortized cost while satisfying the \textit{security requirement} when the protocol terminates.

 \begin{figure}
  \fbox{\parbox[c][][t]{8.4cm}{
   \vspace{2mm}
  \textbf{Input:} $R$, $\lambda$.\\
\textbf{Output:} $(B^{*},T^{*})$.

\vspace{2mm}
For $B$ from 2 to $\lambda$:

\begin{enumerate}
  \item For $T$ from min($\beta, B$) to $+\infty$, find the smallest $T$ that satisfies 
  $\text{Pr}_{success} \leq 2^{-\lambda}$.
  \item
  If $\textit{cost}(B, T, R)<\textit{cost}(B^{*},T^{*}, R)$, set $T^{*}=T$, $B^{*}=B$.
  
\end{enumerate}
\vspace{2mm}
Output $(B^{*},T^{*})$ pair that minimizes the amortized cost function.
  }
  }
        \caption{ \textbf{Protocol $\Pi_{\textit{Search}}$ for searching optimized $B$ and $T$.}}  
            \label{optimizedparameters}
            %\vspace{-2mm}
 \end{figure}

 \vspace{1mm}
\noindent\textbf{(II) Privacy-Preserving Inference Execution.}
In this phase, the client consumes inference services provided by the server by feeding the privately mixed dataset as its input. The server and the client jointly perform  the secure inference computations using the known semi-honest privacy-preserving inference protocols (e.g., Cheetah, C\textsc{ryp}TF\textsc{low}2, etc). For every sample in the mixed dataset, the semi-honest 2PC-based inference protocol is invoked to obtain the inference result.
Finally, the inference results on all samples in the mixed dataset are only revealed to the client. 
At the end of this phase, the correctness of the inference results is not guaranteed and will be checked in the next phase. Please note that \textit{Fusion} provides flexibility in choosing the 2PC-based privacy-preserving inference protocols. Given those are known protocols,  we omitted the details for brevity.

\vspace{1mm}
\noindent\textbf{(III) Model Accuracy and Computation Correctness Verification.}
When obtaining inference results on the mixed dataset, the client checks the \textit{model accuracy} and \textit{computation correctness}. Particularly, the model accuracy $\eta$ is defined as the number of correct inferences on $T$ public samples over $T$. When $\eta$ is greater than a threshold (e.g., 0.95), then \textit{model accuracy} is passed. Similarly, the client verifies the computation correctness by checking the consistency of the inference results on $B$ copies of each query sample.
  If any of the $B$ copies of a query sample are inconsistent, it is considered that the server attempted to deceive the client by giving incorrect inference results. The client accepts the inference results if both checks pass, otherwise, it aborts.

%******************************************
\section{Security Analysis}
\label{securityanalysis}

In this section, we give the security analysis of \textit{Fusion}. In particular, we denote our protocol $\Pi_{{Fusion}}$. 

\vspace{2mm}
\fbox{\parbox[c][][t]{8.2cm}{
\begin{theorem}
 \textit{Assuming the existence of $\Pi_{\mathrm{PPML}}^{\mathrm{semi}}$ that securely achieves inference functionality $\mathcal{F}_{\mathrm{PPML}}^{\mathrm{semi}}$ under the semi-honest security, the protocol $\Pi_{{Fusion}}$ for executing secure inference securely achieves the ideal functionality $\mathcal{F}_{{Fusion}}$ (with abort) against a semi-honest client and a malicious server in the $\mathcal{F}_{\mathrm{PPML}}^{\mathrm{semi}}$-hybrid model.} 
\end{theorem}
}}
\vspace{2mm}

We prove \sys is secure against a semi-honest client and a malicious server in the ideal or real simulation paradigm where the view in both the ideal and real worlds are indistinguishable.
Hence, we need to prove that there exists a simulator in the ideal world that can simulate a view that is indistinguishable from the view in the real world for the adversary who corrupts either the client or the server.

\vspace{1mm}
\noindent \textbf{Compromised Client (CC):}
We first consider the case that the client is corrupted by a semi-honest adversary.
As described in Definition \ref{securitydefinition}, we have to prove that there exists a simulator $\textsf{Sim}_{\mathcal{C}}$ that can simulate the view that is indistinguishable from the view in the real world for the adversary.
Since the client is semi-honest and acts the identical way as the client in $\mathcal{F}_{\mathrm{PPML}}^{\mathrm{semi}}$ except for additional local checks, there exists a simulator $\textsf{Sim}_{\mathcal{C}}$ that can simulate the indistinguishable view by invoking the simulator of $\mathcal{F}_{\mathrm{PPML}}^{\mathrm{semi}}$.

\vspace{1mm}
\noindent \textbf{Compromised Server (CS):}
We then proceed to the case that the server is corrupted by a malicious adversary and construct a simulator $\textsf{Sim}_{\mathcal{S}}$.
When the server is corrupted by a malicious adversary who can deviate from the protocol, the final output of the client may be incorrect and cause abort in both worlds. 
Specifically, we construct a simulator $\textsf{Sim}_{\mathcal{S}}$ that simulates the view in the ideal world.
We will show that the view in the ideal world is statistically indistinguishable from the view in the real world.
In step 2 of $\Pi_{{Fusion}}$ (\autoref{Fusion}), we let simulator  $\textsf{Sim}_{\mathcal{S}}$ invoke the simulator of $\mathcal{F}_{\mathrm{PPML}}^{\mathrm{semi}}$ and output whatever it outputs.
Specifically, these hybrid inference protocols use additive sharing to connect two layers and the intermediate results that the corrupted server obtains are random secret shares that can be simulated by picking uniformly random values.
Next in step 3 of $\Pi_{{Fusion}}$, the simulator  $\textsf{Sim}_{\mathcal{S}}$ works as follows.\looseness=-1

\noindent \textit{(I) CS-Case 1:}
If the adversary does not cheat throughout the protocol, there are two probable results, i.e., (1) the model accuracy fulfills the requirement and the simulator will not send abort to the functionality, and (2) the model accuracy does not meet the requirement and the simulator will send abort to the functionality at the end of the protocol execution.
The simulator is given the corrupted party's input \cite{Lindell17} and it can check the model accuracy, mirroring the behavior of the client.
If the model accuracy does not satisfy the threshold, the simulator will send abort to the functionality while in the real world the client rejects the results and aborts. 
If the model accuracy satisfies the threshold, the functionality will not abort and sends inference results to the client while in the real world the client accepts the results.
The view in both the ideal and the real worlds is identical.

\noindent \textit{(II) CS-Case 2:}
If the adversary cheats by using inconsistent model parameters for multiple query samples or deviating from the protocol, the simulator will detect and send abort to the functionality.
In the real world, if the server cheats as above, it will be caught with overwhelming probability.
The adversary can cheat successfully if all copies of the same query sample obtain incorrect-but-consistent inference results.
The \textit{mix-and-check} method ensures that the cheating probability can be negligibly small when appropriately choosing the parameters $B$ and $T$.\looseness=-1

We begin by defining the following \textit{mix-and-check} game, which is equivalent to our protocol $\Pi_{\textit{Fusion}}$ (\autoref{Fusion}).
The server wins the game if it chooses some of the query samples and provides incorrect inference results for them without being caught by the client.
We need to prove that the probability of the server winning the game is negligible, that is, the server cannot distinguish every two samples in the mixed dataset.
If the game's output is 1, the malicious server wins the game and cheats the client successfully.
The probability $\text{Pr}_{success}$ that the server succeeds in cheating equals $\text{Pr}[\text{Game}(\mathcal{S},\mathcal{C},R,B,T)=1]$.\looseness=-1

\vspace{2mm}
\fbox{\parbox[c][][t]{8.2cm}{
\begin{definition}\label{game}
 \textit{The probability that the server $\mathcal{S}$ wins the game is negligible by choosing appropriate $R$, $B$, and $T$.}
\end{definition}}}
\vspace{2mm}

The game $\text{Game}(\mathcal{S},\mathcal{C},R,B,T)$ proceeds as follows.
(i) The client $\mathcal{C}$ prepares the mixed dataset containing $RB+T$ samples and uses them as inputs for privacy-preserving inference computations.
(ii) The server $\mathcal{S}$ selects $iB$ samples ($i$ is the number of query samples $\mathcal{S}$ chooses to fool) in the mixed dataset and returns incorrect inference results for each of them. 
 (iii) The inference results for the remaining samples are computed correctly using the high-quality model.
 The output of the game is 1 if there are $i$ query samples such that inference results for each of them and corresponding copies are incorrect-but-consistent, while inference results for remaining $R-i$ query samples and their copies are obtained by correctly performing the inference computations using the high-quality model.\looseness=-1
 
\vspace{2mm}
\noindent\textbf{Claim 1:} \textit{If $T\geq B$, then for every adversary $\mathcal{S}$, it holds that
$$ \text{Pr}[\text{Game}(\mathcal{S},\mathcal{C},R,B,T)=1]\leq R \dbinom{R B+T}{B}^{-1}. $$}

 We need to show that for every $1 \leq i \leq R$, 
\begin{equation} 
\dbinom{R}{i}\dbinom{R B+T}{i B}^{-1}\leq R \dbinom{R B+T}{B}^{-1}. 
\end{equation}

At first, it can be observed that when $i=1$, the left side of the inequality equals the right side, and thus the equation holds.
Next, assume that $i\geq2$, it suffices to show that: 
$ \dbinom{R}{i}\dbinom{R B+T}{i B}^{-1}\leq  \dbinom{R B+T}{B}^{-1}. $

\vspace{2mm}
It is equivalent to proving that:
\begin{equation} 
  \dbinom{R}{i} \frac{(iB)!(RB+T-iB)!}{(RB+T)!} \leq \frac{B!(RB+T-B)!}{(RB+T)!},\nonumber
\end{equation}

which can be represented as:

\begin{equation}
\dbinom{R}{i} \frac{(iB)!}{B!} \leq \frac{(RB+T-B)!}{(RB+T-iB)!}. \nonumber
\end{equation}

By multiplying both sides with $\frac{1}{(iB-B)!}$ the above inequality can be transferred to
\begin{equation}
  \dbinom{R}{i} \dbinom{iB}{iB-B} \leq\dbinom{RB+T-B}{iB-B}. \nonumber
\end{equation}

Considering the assumption that $T\geq B$ and thus $
\dbinom{RB}{iB-B}\leq\dbinom{RB+T-B}{iB-B}
$, it suffices to prove that 

\begin{equation}
  \label{final}
  \dbinom{R}{ i} \dbinom{iB} { iB-B} \leq \dbinom{RB}{iB-B}.
\end{equation}

To prove that the above \autoref{final} holds, we can consider the both sides of the inequality as following:
  The left side $\dbinom{R}{ i} \dbinom{iB} { iB-B}$ can represent the process: choosing $i$ query samples among $R$ query samples, then choosing $iB-B$ samples from $iB$ copies of the selected $i$ query samples, and using false model parameters to provide inference results for the $iB-B $ samples.
 The right side $ \dbinom{RB}{iB-B}$ can represent the process: choosing $iB-B$ samples from $RB$ samples. 
The above two processes both end with choosing $iB-B$ samples out of $RB$ samples.
Since there is no restriction on the selection of the right process, the number of choices in the right process is strictly larger than that in the left process.
It is sufficient to conclude that the inequality \autoref{final} holds.

\vspace{2mm}
\noindent\textbf{Claim 2:} \textit{
In real execution, if the parameter $B$ and $T$ are properly chosen as in $\Pi_{\textit{Search}}$, then the client aborts with probability at least $2^{-\lambda}$.}
\vspace{2mm}

When $B$ and $T$ are properly chosen as in $\Pi_{\textit{Search}}$, it ensures that $R \dbinom{R B+T}{B}^{-1} \leq 2^{-\lambda} $, then the probability $ \text{Pr}[\text{Game}(\mathcal{S},\mathcal{C},R,B,T)=1]\leq 2^{-\lambda}$.
That is, if the server cheats in the real protocol, the client will detect and abort with probability at least $1-2^{-\lambda}$, while the simulator will definitely detect the server's cheating behavior and abort in the ideal world.
Thus, the view in both the ideal and real worlds is statistically close.

To conclude, in all cases, the view in both the ideal and real worlds is computationally indistinguishable.
Thus, the protocol $\Pi_{Fusion}$ (\autoref{Fusion}) securely realizes the ideal functionality in the $\mathcal{F}_{\mathrm{PPML}}^{\mathrm{semi}}$-hybrid model against the malicious server.
This completes the proof.

%******************************************
\section{Evaluation}
\label{evaluation}
In this section, we test \sys using multiple experiments. In particular, we first explain the experimental setup (\S\ref{subsec:experimentalsetup}), followed by the experiment results (\S\ref{subsec:experiment}).

\subsection{Experimental Setup.} 
\label{subsec:experimentalsetup}
 
\noindent\textbf{Datasets.} In our study, we used 7 different datasets, and we now explain each of them in greater details: 
\begin{itemize}
\item \textit{TIMIT}~\cite{timitdataset} contains speech data of English recordings from 630 speakers for the development of automatic speech recognition systems.
\item \textit{MNIST}~\cite{MNIST} is a commonly used dataset that contains (28$\times$28) images of handwritten digits between 0 and 9.
\item \textit{\textit{CIFAR-10}}~\cite{cifar10} is a standardized consisting of 60,000 (32$\times$32) color images including 10 classes, e.g., bird, automobile, truck, etc.
\item \textit{BC-TCGA}~\cite{gene} is gene expression data and consists of 17,814 genes and 590 samples (including 61 normal tissue samples and 529 breast cancer tissue samples).
\item \textit{GSE2034}~\cite{gene} is gene expression profiles and includes 12,634 genes and 286 breast cancer samples (including 107 recurrence tumor samples and 179 no recurrence samples).
\item \textit{PneumoniaMNIST}~\cite{medMNISTv2} contains pediatric chest X-Ray images, and the task is a binary-class classification of pneumonia or normal.
\item \textit{DermaMNIST}~\cite{medMNISTv2} consists of 10,015 dermatoscopic images of common pigmented skin lesions that are categorized as 7 different diseases.
\end{itemize}

\noindent\textbf{Environment.} 
We performed experiments on two servers running Ubuntu 16.08 with 2.3 GHz Intel Xeon E5 Broadwell Processors and 244GB RAM.
We set the latency as 40 ms which is higher than the latency between two Amazon EC2 machines located in Ohio and Virginia.
Other experiments were carried on Intel Xeon CPU E5-2630 v3 @ 2.40GHz with 128GB of RAM and Intel Xeon CPU E5-2680 v4 @ 2.40GHz with 256GB of RAM.
The bandwidth between the machines were 382 MBps and 44 MBps, and the echo latency were 0.3 ms and 40 ms in the LAN and the WAN setting respectively.

\noindent\textbf{Availability.} Our implementation is publicly available on GitHub: https://github.com/daisy611/Fusion.

\begin{figure}[!ht]
  \centering
  \includegraphics[width=0.95\linewidth]{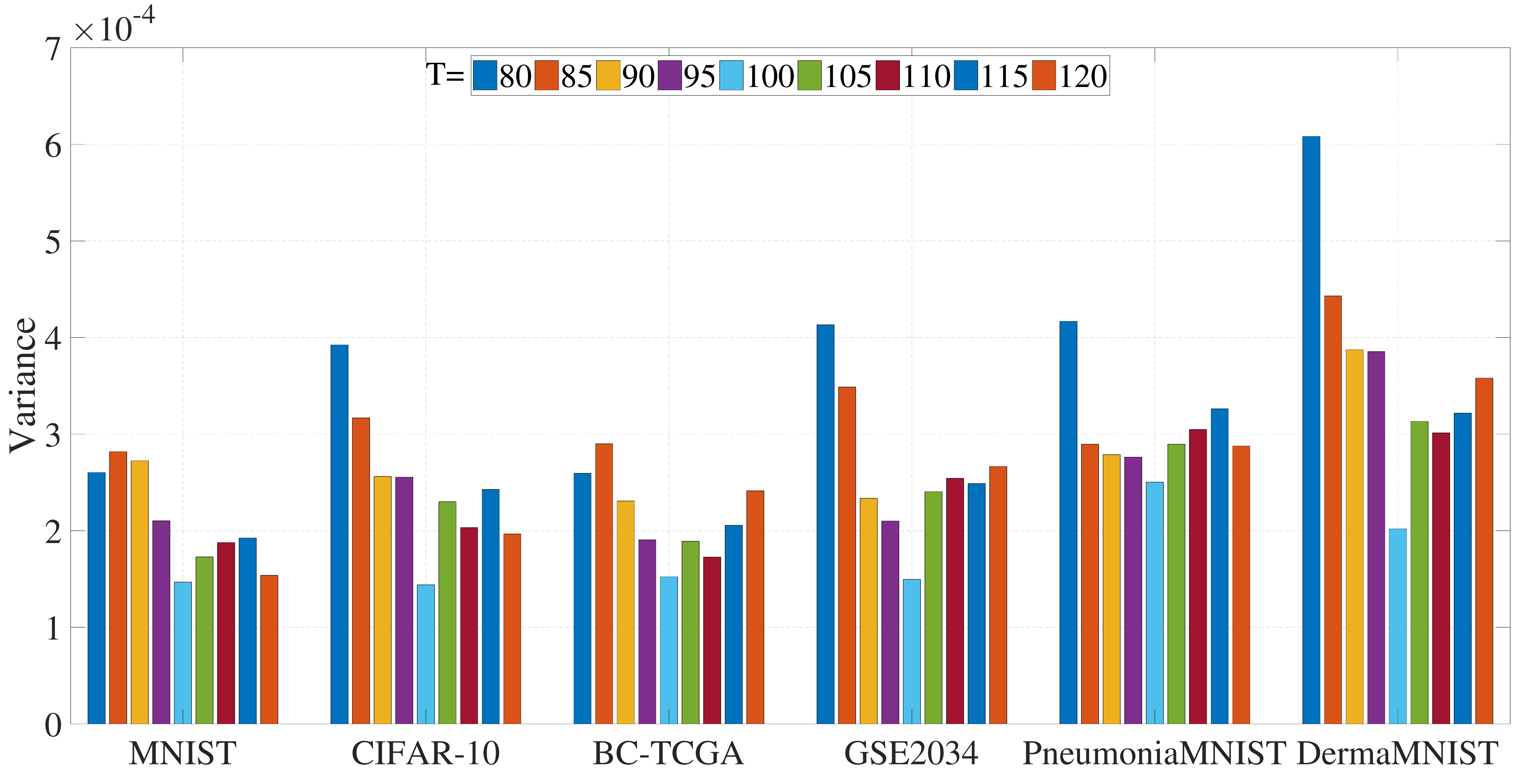}\\
  \caption{The variance of \textit{test accuracy} when the numbers of public samples and datasets vary.}
  \label{variance}
  \vspace{-2mm}
\end{figure}

\noindent\textbf{Selection of Optimal $T$.} 
Our \sys requires a number of public samples $T$ to work.
Therefore, before testing \sys, we would like to explore an optimized number of public samples. To approximate an optimal $T$, we then performed experiments: we first denote the model accuracy that is calculated on a large test dataset as \textit{standard accuracy}, and the accuracy obtained on the $T$ public samples as \textit{test accuracy}.
We test the \textit{test accuracy} of 50 group non-overlapping and uniformly distributed data on six different datasets (e.g., \textit{MNIST}, \textit{CIFAR-10}, \textit{BC-TCGA}, \textit{GSE2034}, \textit{PneumoniaMNIST}, and \textit{DermaMNIST}) for $T=$ 80, 85, 90, 95, 100, 105, 110, 115, and 120.
We calculate the variance of \textit{test accuracy} for different numbers of public samples (by setting the \textit{standard accuracy} as its expected value) and show results in \autoref{variance}.
The empirical results show that when $T=100$ the variance is smallest on all the six datasets.
When $T=100$, the \textit{test accuracy} fluctuates around 2\% above and below the \textit{standard accuracy} (which is acceptable).
Theoretically, based on the analysis in \S\ref{subsec:designsys}, we can notice that $T$ is determined by the \textit{cost requirement} and lower bound of public samples. 
Since the cost per query sample increases as $T$ increases (when $R$ and $B$ are the same), especially the smaller the $R$, the larger the cost per query sample. We therefore empirically selected $T = 100$ for trade-offs. \looseness=-1

\subsection{Experimental Results}
\label{subsec:experiment}

\noindent{\textbf{Comparison with State-of-the-art \textit{LevioSA}}:}
We compare \sys with the state-of-the-art work  maliciously secure neural network inference scheme, i.e., \textit{LevioSA}~\cite{hazay2019leviosa}.
To compare \sys with \textit{LevioSA}, we implemented \sys using Cheetah, and adopted the same settings as \textit{LevioSA}: We employed the same neural network framework as \textit{LevioSA}, which consists of a four-layer deep neural network (DNN) with three hidden, fully connected layers with 2000 neurons, quadratic activations, and the final layer is fully connected with 183 output neurons. We trained the DNN model on the \textit{TIMIT} dataset \cite{timitdataset} as they used.
Since \textit{LevioSA} counted the communication and runtime of inference on 1,845 speech samples, in our experiments, we also set the number of query samples as the same as theirs.
When $R=1,845$, we obtain that optimal parameters $B=5$ and $T=100$ (when setting $\beta=100$) by invoking the search protocol $\Pi_{\textit{Search}}$.
Then we let the client prepare the mixed dataset that contains totally $1,845 * 5 +100 $ samples (1,845 query samples (each with 5 copies) and extra 100 speech samples used as public samples).
We ran experiments by using the trained DNN model to perform inference computations on the  mixed dataset.

\begin{figure}[!ht]
  \centering
  \includegraphics[width=0.65\linewidth]{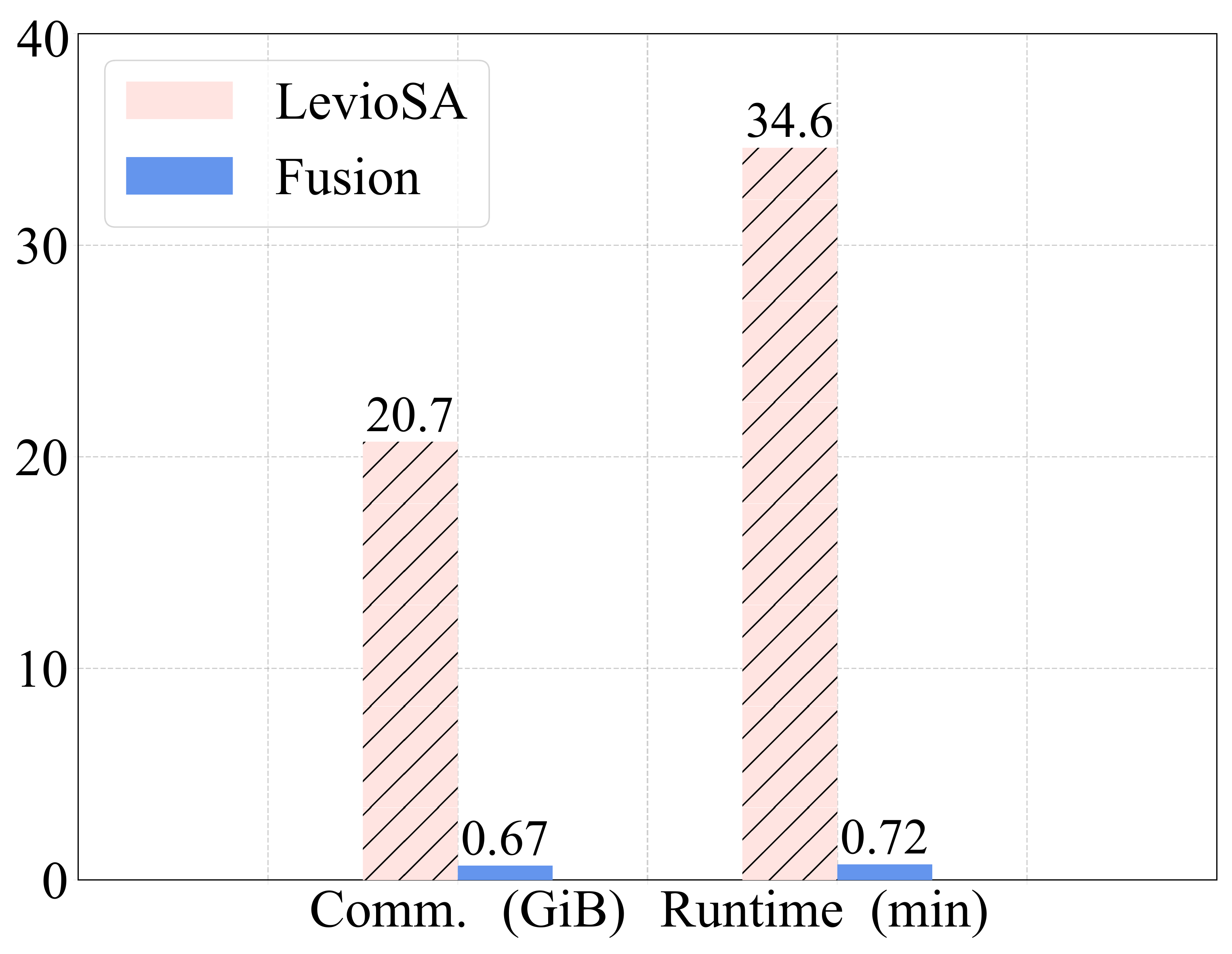}\\
  \caption{Comparison between \textit{LevioSA} and \sys.}
  \label{comparison}
\end{figure}

The experimental results of communication and runtime between \textit{LevioSA} and \sys are shown in \autoref{comparison}.
Our results show that \sys has 48.06$\times$ less runtime and uses 30.90$\times$ less communication compared to \textit{LevioSA}.

\noindent\textbf{Evaluation with Different Underlying Building Blocks:}
The performances of \sys depend on the efficiency of the underlying semi-honest inference protocol (more efficient underlying building block generally improves the performance of \sys), and the choices of $R$, $B$, and $T$ (e.g.,  when $T$ is fixed, the lower $B$ is, the lower the amortized cost of \sys is). As such, to test the performance of \sys,  we change the underlying building blocks (e.g., ABY~\cite{demmler2015aby}, 
 DELPHI~\cite{mishra2020delphi}, C\textsc{ryp}TF\textsc{low}2 ($\text{SCI}_{\text{HE}}$)~\cite{rathee2020cryptflow2}, and Cheetah~\cite{HuangLHD22}), and select different $R$, $B$, and $T$. In particular, since the optimal choice of $B$ to minimize amortized cost is influenced by the number of query samples (e.g., see \autoref{p8}), we need to select various $R$ that lead to different $B$.
To that end, for each $B$ ranging from eight to three, we look for the corresponding $R$ and $T$ that satisfy the \textit{security requirement} $\text{Pr}_{success}\leq 2^{-\lambda}$.
We set the statistical parameter $\lambda = 40$ and empirically choose the least number of query samples $\beta = 100$.
Specifically, for each $B$ from eight to three, we search for the least $R$, and for simplicity, we select $R$ from $2^{x}$ $(x\in \mathbb{Z})$.
The search results are shown in \autoref{parameter}. Having decided $(R,B,T)$, we then conducted experiments on two datasets (\textit{MNIST} and \textit{CIFAR-10}) using multiple building blocks.
Similarly, we used the four-layer DNN network, and implemented \sys  with the settings in~\cite{LeeEMS19} to train models on \textit{MNIST} and \textit{CIFAR-10} datasets respectively.
We then used those models to perform inference computations in different network settings (i.e., LAN and WAN). \looseness=-1

%*********************************************************
  \begin{table}[h]
    \renewcommand{\arraystretch}{1.15}
    \caption{The values of $(R, T)$ with different $B$.}
    \label{parameter}
    \scriptsize
      \centering
      \setlength{\tabcolsep}{1.5mm}{  
    \begin{tabular}{ccccccc}
    \toprule[1.5pt]
    \multicolumn{1}{c}{\textbf{\textit{B}}}    & \textbf{8}   & \textbf{7}                          & \textbf{
6}                          & \textbf{5}                          & \textbf{4}                               & \textbf{3}                               \\ \midrule[0.8pt]   \textbf{(\textit{R, T})}   & ($2^3$, 100)   & ($2^5$, 100) & ($2^7$, 100) & ($2^9$, 100) & ($2^{13}$, 100) & ($2^{19}$, 100)\\ \bottomrule[1.5pt]
    \end{tabular}
    }
    \end{table}
%*********************************************************

The total communication and runtime of \sys when using different $R$ (resulting in different $B$) in the \textit{MNIST} and \textit{CIFAR-10} datasets are shown in \autoref{MNISTLANWAN} and \autoref{CIFARLANWAN} respectively.
It can be observed that \sys can achieve the best efficiency when using Cheetah~\cite{HuangLHD22} as the building block, while when \sys uses ABY~\cite{demmler2015aby} and DELPHI~\cite{mishra2020delphi}, the performance is not as good as that of using others. 
For example, the  Cheetah-based \sys can save around 10$\times$ communication, and is roughly 2.5$\times$ and 4.8$\times$ faster (in the LAN and WAN respectively) compared with those using C\textsc{ryp}TF\textsc{low}2 on both the \textit{MNIST} and \textit{CIFAR-10} datasets. 
When comparing the performance of DELPHI-based \sys,  Cheetah-based \sys uses more than 105$\times$ less communication and is about 22$\times$ faster and 37$\times$ faster (in the LAN and WAN settings respectively) on these two datasets.\looseness=-1

%*********************************************************
  \begin{table*}[!th]
    \caption{Comparison of \sys instantiated using ABY~\cite{demmler2015aby}, DELPHI~\cite{mishra2020delphi}, C\textsc{ryp}TF\textsc{low}2 ($\text{SCI}_{\text{HE}}$)~\cite{rathee2020cryptflow2}, and Cheetah~\cite{HuangLHD22} on the \textit{MNIST} dataset in both LAN and WAN settings. We use Comm. to denote communication and is in MiB, and runtime  is in seconds.}\label{MNISTLANWAN}
    \vspace{-4mm}
    \centering
      \scriptsize
      \setlength{\tabcolsep}{1mm}{
      \begin{tabular}{lrrrrrrrrrrrrr}
      \\
       \toprule[1.5pt]
      \multicolumn{1}{l}{\multirow{3}{*}{\textbf{\# of Samples}}} & \multicolumn{12}{c}{\textbf{\textit{(MNIST)}}}                                                            \\ 
       & \multicolumn{3}{c}{\textbf{ABY}} & \multicolumn{3}{c}{\textbf{DELPHI}} & \multicolumn{3}{c}{\textbf{C\textsc{ryp}TF\textsc{low}2}} & \multicolumn{3}{c}{\textbf{Cheetah}} \\\cmidrule(r){2-4} \cmidrule(r){5-7} \cmidrule(r){8-10} \cmidrule(r){11-13}
      & \textbf{Comm.}              & \textbf{LAN}    & \textbf{WAN}        & \textbf{Comm.}                & \textbf{LAN}     & \textbf{WAN}            & \textbf{Comm. }              & \textbf{LAN}     & \textbf{WAN }            & \textbf{Comm.}               & \textbf{LAN }    &\textbf{ WAN }     \\  \midrule[0.8pt]
         
       164 ($2^3$ * 8 + 100)                   &   27.4949                         & 0.1211         &0.3843               & 20.6144                    & 0.0862        &0.2583               & 1.9984                     & 0.0101       &0.0333               & 0.1914                    & 0.0039 &0.0068
       \\   
     324 ($2^5$ * 7 + 100)                    & 54.4965                         & 0.2407       &0.7386                & 40.7033                    & 0.1711          &0.5128             & 3.9787                     & 0.0201     &0.0676                 & 0.3782                    & 0.0073  &0.0136
        \\
      868 ($2^7$ * 6 + 100)                & 146.0627                        & 0.6393        &1.9728                & 107.6844                   & 0.4619          &1.3667             & 10.5904                    & 0.0541        & 0.1808              & 1.0133                    & 0.0198  &0.0363
        \\
      2,660 ($2^9$ * 5 + 100)                   &441.7787 & 1.9620 & 5.9044  & 335.8445                   & 1.4615  & 4.2086                     & 32.6136                    & 0.1651  &0.5529                    & 3.1048                    & 0.0605 & 0.1105
         \\
     32,868 ($2^{13}$ * 4 + 100)               &5,478.9059                       & 24.3705  &73.9667                     & 4,124.8922                  & 18.1847    &51.7762                  & 400.9943                   & 2.0339   &6.7420                   & 38.3905                   & 0.7429  &1.3678
        \\
     1,572,964 ($2^{19}$ * 3 + 100)               &262,641.3367                     & 1,166.8446      &3,607.8405               & 197,253.8050                & 887.0325    &2,475.5588                  & 19,340.8286                 & 98.3081     &327.7937                & 1,832.7539                 & 35.7586 &  65.8347
                \\\bottomrule[1.5pt]
      \end{tabular}
      }
      \vspace{3mm}
      \end{table*}
%*********************************************************

%*********************************************************
  \begin{table*}[!th]
    \renewcommand{\arraystretch}{1.15}
    \caption{Comparison of \sys instantiated using ABY, DELPHI, C\textsc{ryp}TF\textsc{low}2, and Cheetah on the \textit{CIFAR-10} dataset.}
      \label{CIFARLANWAN}
      \vspace{-4mm}
      \centering
      \scriptsize
      \setlength{\tabcolsep}{1mm}{

      \begin{tabular}{lrrrrrrrrrrrrrr}
      \\
      \toprule[1.5pt]
      \multicolumn{1}{l}{\multirow{3}{*}{\textbf{\# of Samples}}}  & \multicolumn{12}{c}{\textbf{\textit{(CIFAR-10)}}}                                                                              \\ 
        & \multicolumn{3}{c}{\textbf{ABY}} & \multicolumn{3}{c}{\textbf{DELPHI}} & \multicolumn{3}{c}{\textbf{C\textsc{ryp}TF\textsc{low}2}} & \multicolumn{3}{c}{\textbf{Cheetah}} \\\cmidrule(r){2-4} \cmidrule(r){5-7} \cmidrule(r){8-10} \cmidrule(r){11-13}
    & \textbf{Comm.}              & \textbf{LAN}    & \textbf{WAN}        & \textbf{Comm.}                & \textbf{LAN}     & \textbf{WAN}            & \textbf{Comm. }              & \textbf{LAN}     & \textbf{WAN }            & \textbf{Comm.}               & \textbf{LAN }    &\textbf{ WAN }               \\
                                         \midrule[0.8pt]

 164 ($2^3$ * 8 + 100)                  &   33.1801                         & 0.1405                        & 0.4155                     & 25.4172                      & 0.1016                     & 0.2982                       & 2.4810                     & 0.0121                      & 0.0387                     & 0.2350                      & 0.0046                    & 0.0078 
       \\   
      324  ($2^5$ * 7 + 100)                & 65.5365                         & 0.2776                        & 0.8237                     & 50.3487                      & 0.1994                     & 0.5907                       & 4.9215                     & 0.0239                      & 0.0768                     & 0.4637                      & 0.0091                    & 0.0154
        \\
      868 ($2^7$ * 6 + 100)                & 175.8464                        & 0.7456                        & 2.1523                     & 134.9016                     & 0.5372                     & 1.5797                       & 13.1828                    & 0.0639                      & 0.2051                     & 1.2436                      & 0.0243                    & 0.0414 
        \\
       2,660 ($2^9$ * 5 + 100)                &537.8853 & 2.2732 & 6.8505                     & 409.4743                     & 1.6361                     & 4.8425                       & 40.0124                    & 0.1961                      & 0.6345                     & 3.8084                      & 0.0745                    & 0.1261 
         \\
        32,868  ($2^{13}$ * 4 + 100)             &  6,659.7936                       & 28.1152                       & 83.2735                    & 5,140.3375                    & 20.1743                   & 59.4230                      & 497.8308                   & 2.4246                      & 7.7918                     & 47.0841                     & 0.9203                    & 1.5706  
        \\
      1,572,964  ($2^{19}$ * 3 + 100)          &318,618.4824                     & 1,337.5954                     & 4,075.8356                  & 245,896.7141                  & 971.4180                   & 2,854.9131                    & 23,768.0474                 & 115.9847                    & 374.3836                   & 2,256.4829                   & 44.1961                   & 75.1037
                \\\bottomrule[1.5pt]
      \end{tabular}
      }
      \vspace{-2mm}
      \end{table*}
%*********************************************************

\vspace{1mm}
\noindent\textbf{Comparison with Semi-honest Solutions:}
We would like to understand whether \sys can be more efficient when compared with semi-honest inference protocols (i.e., ABY~\cite{demmler2015aby}, DELPHI~\cite{mishra2020delphi}, and C\textsc{ryp}TF\textsc{low}2 ($\text{SCI}_{\text{HE}}$)~\cite{rathee2020cryptflow2}). To that end,  we conducted experiments using the DNN network  on both \textit{MNIST} and \textit{CIFAR-10} datasets and both network settings (the LAN and WAN settings).
We change $R$ as $2^{3}$, $2^{5}$, $2^{7}$, $2^{9}$, $2^{13}$, and $2^{19}$, and the corresponding $B$ for each $R$ ranges from 8 to 3 (with fixed $T=100$). \looseness=-1

In \autoref{differentT}, we compare the communication and runtime per query sample of \sys  with those of ABY~\cite{demmler2015aby}, DELPHI~\cite{mishra2020delphi}, and C\textsc{ryp}TF\textsc{low}2~\cite{rathee2020cryptflow2}.
Note that the amortized cost of \sys decreases as $B$ decreases, and greater performance improvements will be achieved.
The experimental results imply that \sys is more efficient than ABY, DELPHI, and C\textsc{ryp}TF\textsc{low}2 in terms of communication when $R \geq 2^{5}$, $B \leq 7$, and $T=100$.
For instance, when using the \textit{MNIST} dataset and setting $R=2^{5}$, $B=7$, $T=100$, the communication cost of \sys is 14.13$\times$, 10.61$\times$, 1.04$\times$ lower than those of ABY, DELPHI, C\textsc{ryp}TF\textsc{low}2, respectively.
When $R \geq 2^{5}$ and corresponding $B \leq 7$, \sys is faster than ABY and DELPHI in both the LAN and WAN settings.
For example, when $R=2^{9}$, $B=5$, and $T=100$, \sys is about 10.52$\times$ and 7.37$\times$ faster than ABY and DELPHI respectively on the \textit{CIFAR-10} dataset in the WAN setting.
\sys uses about 2.62$\times$ less communication, and is 1.25$\times$ faster than C\textsc{ryp}TF\textsc{low}2 on \textit{MNIST} dataset in the WAN setting when $R = 2^{13}$ and $B=4$. \looseness=-1

%*********************************************************
\setlength{\tabcolsep}{1.5mm}{
  \begin{table}[h]
    \renewcommand{\arraystretch}{1.15}
    \caption{Performance of Cheetah-based \sys (with different ($R, B, T$)), and comparison with semi-honest inference protocols. Comm. is in KiB, and runtime is in \textmu s.}
    \label{differentT}
    \centering
    \scriptsize
     \setlength{\tabcolsep}{1mm}
    
    \begin{tabular}{lrrrrrrr}
    \toprule[1.5pt]
    \multicolumn{2}{c}{\multirow{2}{*}{\textbf{Scheme}}}
       & \multicolumn{3}{c}{\textbf{\textit{MNIST}}} & \multicolumn{3}{c}{\textbf{\textit{CIFAR-10}}}  \\   \cmidrule(r){3-5} \cmidrule(r){6-8}
      & & \textbf{Comm.}                & \textbf{LAN}     & \textbf{WAN }       & \textbf{Comm. }                &\textbf{ LAN}     & \textbf{WAN}    \\
       \midrule[0.8pt]
       \multirow{5}{*}{\begin{tabular}[c]{@{}l@{}} \rotatebox[origin=c]{90}{\textbf{Fusion}}\end{tabular}}
    &($2^3$, 8, 100)                     & 24.499     & 487.500       & 850.000  & 30.080       & 575.000         & 975.000        \\ 
    
     & ($2^5$, 7, 100)                
     & 12.102     & 228.125     & 425.000       & 14.838     & 284.375     & 481.250      \\
    
      & ($2^7$, 6, 100) 
    & 8.106      & 154.688    & 283.594  &9.949      & 189.844   & 323.438  
    \\ 
          & ($2^9$, 5, 100) 
        & 6.210      & 118.164 & 215.820 & 7.617      & 145.508 & 246.289 \\  
    & ($2^{13}$, 4, 100) 
    & 4.799   & 90.686 & 166.968 &  5.886   & 112.341 & 191.724 \\  
   & ($2^{19}$, 3, 100) 
   & 3.580 & 68.204 & 125.570  & 4.407 & 84.297 & 143.249 \\   
  \cmidrule{1-8}
         \multicolumn{2}{r}{\textbf{C\textsc{ryp}TF\textsc{low}2}~\cite{rathee2020cryptflow2}}
  & 12.591                  & 62.499 & 208.392 &15.473                                 & 73.736                                & 238.012 \\  
\multicolumn{2}{r}{\textbf{DELPHI}~\cite{mishra2020delphi}}  & 128.412  & 563.924  &  1573.818  & 160.079 & 617.572     & 1814.989    \\
\multicolumn{2}{r}{\textbf{ABY}~\cite{demmler2015aby}}   &  170.980    &    741.813   & 2293.657 &207.421  & 850.366  &  2591.182      \\
    \bottomrule[1.5pt]
    \end{tabular}
  
    \end{table}
    }
%*********************************************************

\vspace{1mm}
\noindent \textbf{Evaluation on Medical Datasets:} Since medical data is high-sensitive and has strong privacy preservation requirements, medical data analysis is one of the most important applications of secure inference.
To show the applicability of \sys in real-world medical datasets, we evaluated experiments on four publicly available healthcare datasets.
 We used the same four-layer DNN network  with settings~\cite{LeeEMS19} to train these models and perform inference. 
 \autoref{medicaldataset} shows the experimental results on four medical datasets \textit{BC-TCGA}~\cite{gene}, \textit{GSE2034}~\cite{gene}, \textit{PneumoniaMNIST}~\cite{medMNISTv2}, and \textit{DermaMNIST}~\cite{medMNISTv2}.
Take the \textit{BC-TCGA} dataset as an example, it consists of 17,814 genes (features), and costs 14.579 ms and 0.469 MiB respectively in maliciously secure inference per query sample. As such, the runtime cost and communication cost are acceptable. \looseness=-1 

%*********************************************************
 
  \begin{table}[h]
    \renewcommand{\arraystretch}{1.1}
    \caption{Performance on medical datasets. It shows amortized communication (MiB) and runtime (ms) per query sample when $R=2^{13}$, $B=4$, and $T=100$.}
    \label{medicaldataset}
    \centering
    \scriptsize
 \setlength{\tabcolsep}{5mm}

    \begin{tabular}{lrrr}
    \toprule[1.5pt]
    \multicolumn{1}{l}{\textbf{Dataset}} & \multicolumn{1}{c}{\textbf{Comm}.} & \textbf{LAN} & \textbf{WAN} \\ \midrule[0.65pt]
    \textit{BC-TCGA} \cite{gene}                  & 0.469                       & 8.710   &14.579 \\ 
    \textit{GSE2034} \cite{gene}                 & 0.297                       & 6.628  &11.342  \\ 
    \textit{PneumoniaMNIST} \cite{medMNISTv2}
          & 0.694                    & 26.864  &44.931 \\ 
          \textit{DermaMNIST} \cite{medMNISTv2}
          & 0.034                   & 2.353 & 3.928  \\    

    \bottomrule[1.5pt]
    \end{tabular}
    
    \vspace{-2mm}
    \end{table}
%*********************************************************

\vspace{1mm}
\noindent \textbf{Evaluation on Practical DNN ResNet50:}
To demonstrate the scalability of \sys, we also conducted evaluations of maliciously secure inference on ImageNet-scale deep neural networks (DNN) ResNet50 \cite{HeZRS16}.
We trained a model on ResNet50 using an image dataset \cite{imagenet}, and used it to perform inference computation.
Since $R$ affects the amortized cost of \sys, we choose $R$ as $2^{3}$, $2^{5}$, $2^{7}$, and $2^{9}$, and the corresponding number of copies for each query sample varies from 8 to 5 (when $T=100$).
We also conducted experiments by using the semi-honest C\textsc{ryp}TF\textsc{low}2 protocol and compared it with \sys.
 \autoref{resnetimagenet} shows the communication and runtime per query sample.
 It can be observed that when $R\geq 2^{5}$ and corresponding $B \leq 7$, \sys is more efficient than the semi-honest C\textsc{ryp}TF\textsc{low}2 ($\text{SCI}_{\text{HE}}$) in terms of communication.
For example, when $R=2^{9}$ and $B=5$, \sys costs 1.30$\times$ runtime and is 1.18$\times$ faster in the LAN setting and the WAN setting respectively, and has 2.64$\times$ less communication than that of C\textsc{ryp}TF\textsc{low}2 ($\text{SCI}_{\text{HE}}$).

%*********************************************************
\setlength{\tabcolsep}{4mm}{
  \begin{table}[h]
    \caption{Performance on ResNet50 and ImageNet-scale benchmarks. It shows communication cost (GiB) and runtime (minutes) per query sample with different ($R, B, T$).}
    \label{resnetimagenet}
    \centering
    \scriptsize
    \renewcommand{\arraystretch}{1}
    \begin{tabular}{crrr}
    \toprule[1.5pt]
    \multicolumn{1}{c}{\multirow{1}{*}{\textbf{Scheme}}} & \textbf{Comm}. & \textbf{LAN} & \textbf{WAN} \\
    \midrule[0.8pt]
    \textbf{\sys} ($2^3$, 8, 100)                     & 39.921                       & 20.410   &34.241 \\ 
    \textbf{\sys} ($2^5$, 7, 100)                   & 19.714                       & 10.082  &16.912  \\ 
       \textbf{\sys}  ($2^7$, 6, 100) 
          & 13.205                     & 6.750  &11.326 \\ 
          \textbf{\sys} ($2^9$, 5, 100) 
          & 10.117                     & 5.173 &8.678  \\    \cmidrule{1-4}
          \textbf{C\textsc{ryp}TF\textsc{low}2} \cite{rathee2020cryptflow2} 
 ($\text{SCI}_{\text{HE}}$)
          & 26.742                     & 3.988  & 10.204 \\  
          \textbf{C\textsc{ryp}TF\textsc{low}2} \cite{rathee2020cryptflow2} ($\text{SCI}_{\text{OT}}$)
          & 281.497                     & 4.795  & 39.466 \\  
    \bottomrule[1.5pt]
    \end{tabular}
    \end{table}
    }
%*********************************************************

\section{Discussion}

\subsection{Defense against Model Extraction Attacks}
In MLaaS, a semi-honest client who correctly follows the protocol still has some other approaches to steal the server's model, and the black-box model extraction attack~\cite{TramerZJRR16, PalGSKSG20, CarliniJM20, JagielskiCBKP20} is one of them. In this type of attack, the client with black-box access keeps querying the model to  extract an equivalent ML model.  
Since this type of attack is getting more attention recently, and the threat model of this attack is consistent with ours (i.e., the client is semi-honest),  in this section, we would like to discuss the scalability of \sys (e.g., how hard or easy to integrate prior defenses into \sys) in defending against model extraction attacks. As such, the researchers who would like to use our \sys and also have the requirements of defending against model extraction attacks can benefit from our paper in a timely manner. 

While it is true that there are already numerous defense measures that can work against model extraction attacks (e.g.,~\cite{jia2021entangled,xian2022framework,yan2021monitoring}), intuitively, we can directly use any of those off-the-shelf solutions. However, we would like to note there should be at least two criteria that need to be fulfilled. First, \sys can verify model accuracy (R1), computation correctness (R2), and preserve privacy (R3). The integrated solution should not introduce weaknesses to break those guarantees. Second, \sys itself is cryptographic-friendly, and as such, we should not introduce heavy cryptographic-based solutions for performance concerns.
Fortunately, we adopt a passive defense method~\cite{LeeEMS19} that satisfies all those design criteria by  perturbing the model's   activation layer to change the output probabilities. As the solution directly makes changes to the neural network, it will not introduce extra overhead on \sys, and allows \sys to achieve the three requirements even when the solution is adopted. Consequently, by integrating the solution, \textit{Fusion} is defense-effective against the model extraction attacks (e.g., the accuracy of the client's stolen model drops by up to 42.7\% compared with when not using defense) meanwhile maintaining the {\textit{utility}} (e.g., reducing the model accuracy with no defense by 1.75\%). Moreover, the presence of defense only causes low extra costs, accounting for less than 1\% of the overall runtime and communication. We have detailed design and evaluation in Appendix \S\ref{appendix:modelextraction} for readers of interest.

\subsection{Selection of Underlying Semi-honest Inference Protocols}

The prominent contribution of \sys lies in providing a flexible and efficient compiler for secure inference, which achieves verification of model accuracy and computation correctness effectively. As a general compiler, \sys offers flexibility in selecting semi-honest inference protocols as underlying building blocks to instantiate \sys. Specifically, these semi-honest inference protocols may utilize various sub-protocols such as secret sharing protocols, HE, GC, OT, etc. As such, the malicious server plays distinct roles within different sub-protocols, e.g., a computing party in HE, a garbler in GC, a receiver in OT, etc. The server may perform various malicious behaviors according to specific sub-protocols and corresponding roles. When the underlying semi-honest inference protocol is very complicated and uses many types of sub-protocols, it becomes more challenging to detect the malicious server's misbehavior and prevent privacy leakage. When the malicious server possesses stronger ability in some parts of the sub-protocols, it can lead to unforeseen complications. Therefore, a judicious selection of the underlying building blocks is important to mitigate unpredictable privacy issues. 

To analyze the selection of underlying semi-honest inference protocols for instantiating \sys, we roughly classify secure inference protocols into three categories based on the cryptographic techniques utilized: ($i$) secret sharing based protocols that use Arithmetic sharing and Boolean sharing based protocols, e.g., C\textsc{ryp}TF\textsc{low}~\cite{kumar2020cryptflow}. ($ii$) hybrid protocols that use HE and GC to perform linear and non-linear layers respectively, e.g., DELPHI~\cite{mishra2020delphi}. ($iii$) hybrid protocols that use HE and Millionaire protocol to perform linear and non-linear layers respectively, e.g., C\textsc{ryp}TF\textsc{low}2~\cite{rathee2020cryptflow2} and Cheetah~\cite{HuangLHD22}. C\textsc{ryp}TF\textsc{low}2 and Cheetah can improve the efficiency of non-linear layers by using the Millionaire protocol than using GC to perform non-linear layers like DELPHI. In addition, when using GC in DELPHI, the malicious server can act as a malicious GC garbler or malicious OT receiver, and tries to steal additional information by misbehaving in the protocols. Although these problems may be avoided by assigning the malicious server the role of a GC evaluator or OT sender, we can alternatively choose other types of protocols that do not rely on GC to perform non-linear layers.

To avoid these possible problems, we suggest choosing a secret sharing based comparison protocol to implement non-linear layers. When using secret sharing based protocols (e.g., C\textsc{ryp}TF\textsc{low}) to achieve both linear and non-linear layers, the intermediate results a malicious server obtains are random secret shares. Additionally, hybrid protocols using the Millionaires protocol to perform non-linear layers have a similar property. For instance, Cheetah and C\textsc{ryp}TF\textsc{low}2 perform non-linear layer by using secret sharing based Millionaire protocol, which involves 1-out-of-n OT and secret sharing based AND operation. If the malicious server acts as the sender, a malicious sender can only cause incorrect results. Else if the server acts as the receiver, even if a malicious receiver can obtain the sender's all messages, it cannot reveal the sender's input as the messages are inputs masked by randomly sampled values. By preventing the malicious server from acting as the receiver, we avoid the scenarios where it possesses more power to conduct malicious behaviors and extract information. For other unforeseen cases, careful selection of sub-protocols or suitable roles for the server within the sub-protocols can be explored as potential solutions. Overall, we recommend choosing secret sharing based inference protocols to instantiate \sys.\looseness=-1

%******************************************
\section{Related Work}
\label{relatedwork}
 This paper studies secure inference protocols.
We categorize existing works into outsourcing scenarios (\S\ref{outsourcing}) and non-outsourcing scenarios (\S\ref{nonoutsourcing}). Since \sys works in non-outsourcing scenarios,  we particularly focus on the later case. \looseness=-1\vspace{-1mm}

\subsection{Secure Inference Protocols in Outsourcing Scenarios}
\label{outsourcing}

In an outsourcing scenario, some servers are responsible for performing the secure inference protocol, and the client does not participate in the inference computations.
Typically, the client provides query samples to and receives inference results from the servers such that the servers cannot know both the client's inputs and inference results. Based on the adopted threat model, the secure inference protocols in the outsourcing scenario can further have two types: 

\vspace{1.5mm}
\noindent\textbf{\textit{Semi-honest Security:}} ABY 1.0\&2.0 \cite{demmler2015aby,patra2021aby2} are mixed 2PC protocols that consider the semi-honest security and provide secure conversions between Arithmetic (A), Boolean, and Yao (Y) sharing.
% These schemes make efforts to reduce the communication rounds or total communication, e.g., making the communication of the dot product to be independent of the vector size~\cite{patra2021aby2}.
Some schemes~\cite{jiang2018secure, chen2019efficient} adopt HE to perform secure outsourced neural network inference and make efforts to improve the performance.

\vspace{1.5mm}
\noindent\textbf{\textit{Malicious Security in Honest Majority Setting:}}
Some schemes~\cite{mohassel2018aby3, rachuri2019trident,patra2020blaze, dalskov2021fantastic, koti2021swift} achieve honest-majority malicious security (i.e., the number of corrupted parties is less than half of the total number of computing parties involved in MPC).
These protocols need three~\cite{mohassel2018aby3, rachuri2019trident, patra2020blaze, koti2021swift} or four~\cite{dalskov2021fantastic} non-colluding servers to perform secure inference protocols.
The aim of these schemes is to achieve better efficiency by various means (e.g., by enabling the communication of dot products in the online phase is independent of the vector size~\cite{patra2020blaze}) or achieve stronger security (e.g., output delivery guarantee by using the property of honest-majority setting~\cite{koti2021swift, dalskov2021fantastic}).

The major difference between \sys and those schemes is the scenarios: those schemes all work in the outsourcing scenario, while \sys work in the non-outsourcing scenario. Moreover, our \sys additionally supports a stronger security level (i.e., malicious security in the dishonest majority).
Moreover, our scheme can ensure the computation correctness and verification of model accuracy in the presence of a malicious server.\looseness=-1

\newcommand\featuretext[1]{
  \llap{\vrule width.35pt height2pt depth2.5pt\kern1pt}%
  \rlap{\rotatebox{40}{\textbf{#1}}}%
}

\begin{table}[t]
  \caption{Comparison of \sys with others. \faStarHalf ~represents ``semi-honest''. \faStarHalfO~represents ``malicious security with honest majority''. \faStar~represents ``malicious security with dishonest majority''. }
\label{tab:comparison}
\centering
\scriptsize
\renewcommand{\arraystretch}{1}
%\resizebox{3in}{!}{
\setlength{\tabcolsep}{5pt}
 \vspace{-0.1in}
\begin{tabular}{llccccccccccc} 

    % Table header %
	\textbf{Schemes}  &
%	\featuretext{w/o both BLE \& BT Packets} &
	\featuretext{Threat Model} &
	\featuretext{Methodology} &
	\featuretext{No Extra Hardware} &
	\featuretext{Non-outsourcing?} &
    \featuretext{R1.Model Accuracy}& 
    \featuretext{R2.Correctness }&
	\featuretext{R3. Both Privacy } &

	\\
	\toprule[1.5pt]
	
	% Table body
	% \tickYes \tickNo
  Jiang et al.~\cite{jiang2018secure}    &   \faStarHalf  & HE &   \cmark  & \xmark & \xmark & \cmark & \cmark \\
  Chen et.al.~\cite{chen2019efficient} & \faStarHalf  &  HE &  \cmark  & \xmark  &  \xmark &  \cmark & \cmark  \\
   ABY~\cite{demmler2015aby} & \faStarHalf &  SS+GC &  \cmark  & \xmark  &  \xmark &  \cmark & \cmark   \\
ABY2.0~\cite{patra2021aby2} & \faStarHalf &  SS &  \cmark & \xmark  & \xmark &  \cmark & \cmark   \\
  ABY3~\cite{mohassel2018aby3} &  \faStarHalfO & SS+GC & \cmark  & \xmark  &  \xmark &  \cmark & \cmark   \\
  Trident~\cite{rachuri2019trident}  &  \faStarHalfO &  SS &  \cmark  & \xmark  &  \xmark &  \cmark & \cmark   \\
  Blaze~\cite{patra2020blaze}  &  \faStarHalfO &  SS &  \cmark  & \xmark  &  \xmark &  \cmark & \cmark   \\
  FantasticFour~\cite{dalskov2021fantastic}  &  \faStarHalfO &  SS &  \cmark  & \xmark  &  \xmark &  \cmark & \cmark   \\
  SWIFT~\cite{koti2021swift}   &  \faStarHalfO &  SS &  \cmark  & \xmark  &  \xmark &  \cmark & \cmark   \\
MiniONN~\cite{liu2017oblivious} & \faStarHalf  &  HE+GC+SS &  \cmark & \cmark  &  \xmark &  \cmark & \cmark   \\
GAZELLE~\cite{juvekar2018gazelle} & \faStarHalf  &  HE+GC+SS &  \cmark & \cmark  &  \xmark &  \cmark & \cmark   \\

XONN~\cite{riazi2019xonn} & \faStarHalf  &  GC+SS &  \cmark & \cmark  &  \xmark &  \cmark & \cmark   \\
DELPHI~\cite{mishra2020delphi} & \faStarHalf  &  HE+GC+SS &  \cmark & \cmark  &  \xmark &  \cmark & \cmark   \\
SiRNN~\cite{rathee2021sirnn} & \faStarHalf  &  SS &  \cmark & \cmark  &  \xmark &  \cmark & \cmark   \\
C\textsc{ryp}TF\textsc{low}2~\cite{rathee2020cryptflow2} & \faStarHalf  &  HE+SS+OT &  \cmark & \cmark  &  \xmark &  \cmark & \cmark   \\
Cheetah~\cite{HuangLHD22} & \faStarHalf  &  HE+SS+VOLE &  \cmark & \cmark  &  \xmark &  \cmark & \cmark   \\
% Safetynets~\cite{ghodsi2017safetynets} & \faStar  &  ZK &  \xmark & \xmark  &  \xmark &  \cmark & \xmark   \\
% vcnn~\cite{lee2020vcnn} & \faStar  &  ZK &  \xmark & \xmark  &  \xmark &  \cmark & \xmark   \\
Veriml~\cite{zhao2021veriml} & \faStar  &  ZK &  \cmark & \cmark &  \cmark &  \cmark & \xmark   \\
% ZEN~\cite{feng2021zen}& \faStar  &  ZK &  \xmark & \xmark  &  \xmark &  \cmark & \xmark   \\
zkcnn~\cite{liuzkcnn}& \faStar  &  ZK &  \cmark & \cmark  &  \cmark &  \cmark & \xmark   \\
Mystique~\cite{weng2021mystique}& \faStar  &  ZK &  \cmark & \cmark  &  \cmark &  \cmark & \xmark   \\
C\textsc{ryp}TF\textsc{low}~\cite{kumar2020cryptflow} 
& \faStar &  SS &  \xmark & \xmark  &  \xmark &  \cmark & \cmark   \\
 \textit{LevioSA}~\cite{hazay2019leviosa} 
& \faStar &  SS &  \cmark & \cmark  &  \xmark &  \cmark & \cmark   \\  \cmidrule{1-8}
\sys & \faStar  &  Mix\&Check & \cmark & \cmark  &  \cmark &  \cmark & \cmark  \\ \bottomrule[1.5pt]
\end{tabular}
%}

% \vspace{-3mm}
\end{table}

\subsection{Secure Inference Protocols in Non-outsourcing Scenarios}\label{nonoutsourcing}
In the non-outsourcing scenario, the client participates in performing the 2PC-based secure inference protocol by interacting with the server and finally obtains the inference results. Similarly, there are also two types correspondingly: 
 
 \vspace{1.5mm}
\noindent\textbf{\textit{Semi-honest Security:}}
Due to the efficiency bottleneck, most of the existing secure inference protocols~\cite{liu2017oblivious, juvekar2018gazelle, chen2019efficient, riazi2019xonn, mishra2020delphi, rathee2020cryptflow2,rathee2021sirnn} consider semi-honest security (e.g., assuming both the server and client honestly follow the protocol), and focus on improving efficiency to enable practical applications.
Specifically, many solutions use a hybrid cryptographic protocol (e.g., adopting different cryptographic techniques for linear and non-linear layers respectively) to achieve better efficiency.
% As shown in \autoref{tab:comparison}, when compared with those protocols, our scheme considers a stronger security model (malicious security), and we additionally achieve the verification of model accuracy.\looseness=-1
 
 \vspace{1.5mm}
\noindent\textbf{\textit{Malicious Security:}}
Recently, some works begin to recognize the significance of security against a malicious server in MLaaS, and try to present solutions for achieving computation correctness in the presence of a malicious adversary.

\begin{itemize}
\item 
\textbf{ZK Proof-based Protocols:}
There are some schemes~\cite{zhao2021veriml, liuzkcnn, weng2021mystique} that adopt zero-knowledge proofs to achieve verifiable inference computations.
Verifiable inference services based on ZK proofs allow one party with a secret witness to prove some statement (e.g., the inference results are correctly produced by the expected model) without revealing private information.
Schemes based on ZK proofs have the advantage of being publicly verifiable.
It usually assumes, however, that the client's input is visible to the server.
As a consequence, it is unable to apply to scenarios where both model and client's inputs are required to be protected, while our scheme preserves the privacy of both the server and client's inputs.

\vspace{1mm}
\item 
\textbf{TEE-based Protocols:}
C\textsc{ryp}TF\textsc{low} \cite{kumar2020cryptflow} presents a novel technique, called Aramis, that takes any semi-honest secure MPC protocol for computation and converts it into a malicious secure MPC protocol by using hardware Intel SGX.
Our scheme is a purely cryptographic protocol and does not require the trust in hardware.

\vspace{1mm}
\item \textbf{MPC-based Protocols:}
If there exists a malicious party in 2PC computation, it definitely has to achieve malicious security with the dishonest majority (meaning the number of corrupted parties is greater than or equal to half of the total number of parties involved in MPC).
Naturally, MPC protocols that consider malicious security in the dishonest majority are relatively complicated problems compared with those achieve malicious security in the honest majority setting or semi-honest security.
Since the malicious security in the honest majority setting (a weaker security model) naturally bring dramatic performance improvements and is generally used in an outsourcing scenario, it is unfair to compare them~\cite{rachuri2019trident,patra2020blaze, dalskov2021fantastic, koti2021swift} with \sys which achieves malicious security in the dishonest majority.\looseness=-1
\end{itemize}

As shown in \autoref{tab:comparison},
our scheme is the only scheme that satisfies all the three requirements (i.e., R1, R2 and R3) that proposed in this paper at the same time, while other works have their own different focuses. In particular, our work considers the verification of the inputs (R1), and currently, only a few works have considered it. When compared with those works that fulfilled R1, our work does not use ZK proofs, which is more computational friendly. We would like to note that the most similar work to ours is \textit{LevioSA} \cite{hazay2019leviosa} which achieves maliciously secure 2PC arithmetic computation in the dishonest majority setting and applies it to neural network classification.
% LevioSA works on Arithmetic circuits and the gates perform addition, subtraction, and multiplication operations over $\mathbb{F}$.
It proposes a passive-to-active oblivious linear function evaluation (OLE) compiler by following the high-level approach of IPS compiler \cite{ishai2009secure,lindell2011ips}. Different from their work, our \sys additionally achieves the verification of model accuracy.

%Although they achieve privacy preservation of both parties and computation correctness in the presence of a malicious adversary, they do not consider verification of inputs, and their solution is also not cheap.

%-------------------------------------------------------------------------------

%******************************************
\section{Conclusion}
\label{conclusion}
We propose \textit{Fusion}, which can achieve security requirements including privacy-preserving, verification of model accuracy and computation correctness. 
\textit{Fusion} can be used as a general compiler by converting a semi-honest inference protocol (e.g., Cheetah, DELPHI) into a maliciously secure one and thus can benefit from the state-of-the-art efficient 2PC-based inference scheme. {Our evaluation shows that \sys is 48.06$\times$ faster and has 30.90$\times$ less communication than \textit{LevioSA} (the state-of-the-art maliciously secure inference protocol).
Moreover, evaluation on the practical ImageNet-scale ResNet50 model indicates that \sys can be more efficient than semi-honest inference protocols.}

%\textit{Fusion} maintains the server’s model \textit{utility} while mitigating the model extraction attacks by decreasing the accuracy of the client’s stolen model.
%In the future, we will focus on considering a stronger threat model that the client may give false reports about the inference results, and provide solutions for their reliability.

\section*{Acknowledgment}
We thank the anonymous reviewers for their invaluable comments to improve our paper.
Jian Weng was supported by National Natural Science Foundation of China under Grant No. 61825203, Major Program of Guangdong Basic and Applied Research Project under Grant No. 2019B030302008, National Key Research and Development Plan of China under Grant No. 2020YFB1005600, Guangdong Provincial Science and Technology Project under Grant No. 2021A0505030033, National Joint Engineering Research Center of Network Security Detection and Protection Technology, and Guangdong Key Laboratory of Data Security and Privacy Preserving.
Anjia Yang was partially supported by the National Key R\&D Program of China (2021ZD0112802), the Key-Area Research and Development Program of Guangdong Province(2020B0101090004, 2020B0101360001), the National Natural Science Foundation of China (62072215).

\bibliographystyle{plain}
\bibliography{fusion}

\begin{appendices}
\section{}
% \section{Capabilities of Defending Against Model Extraction Attacks}
\label{appendix:modelextraction}

\subsection{Defense against Model Extraction Attacks}\label{defensemethod}
Existing defense strategies \cite{JuutiSMA19, JiaCCP21, LeeEMS19, ChengLZZ20, OrekondySF20, LehmkuhlMSP21} can be divided into two types, i.e., detecting anomalous queries or analyzing query patterns \cite{JuutiSMA19}, or using perturbation to make it resilient to the model extraction attacks.
The former type makes strong assumptions on the attacker's query distributions and requires query pattern analysis on the client's query samples.
Since the query samples and the output labels (inference results) are private to the server, it is tricky to defend against the client by detecting the client's query patterns \cite{OrekondySF20} or adding perturbation according to specific inference results \cite{LeeEMS19}.
The latter type can be implemented by the server before the inference service, allowing it to collaborate with privacy-preserving inference protocols without compromising the client's input privacy.
These strategies, on the other hand, mitigate model extraction attacks by sacrificing model accuracy, necessitating a trade-off between model utility and defense effectiveness.

Since the query samples and the final inference results are protected from the server in secure inference, the passive defense methods, without requiring additional knowledge about the client's query samples or output labels, are well-compatible with our scheme.
We adopt a passive defense method proposed in \cite{LeeEMS19} because it mitigates the model extraction attacks by perturbing the server's model network.
As the attacker essentially approximates the loss hypersurface to find parameters for the stolen model with the minimum loss value, the defense method leverages a new activation layer that manipulates the estimated loss surface by adding a small controllable perturbation and thus maximizes the loss of the stolen model while preserving the accuracy of the original model.
Specifically, this defense method only perturbs the model's final activation layer (e.g., softmax) that maps a vector to a number of probability values.

The \textit{Reverse Sigmoid} perturbation $r(y_{j}^{i})$ is as follows.
\begin{equation}
  r(y_{j}^{i}) \approx \beta (s(\gamma s^{-1}(y_{j}^{i}))-1/2),
\end{equation}
where $y_{j}^{i}$ represents $j$-th dimension of probability vector $y^i$ for sample $x_i$, $s(\cdot)$ is a sigmoid function, $\gamma$ is a positive dataset and model specific convergence parameter, and $\beta$ is a positive magnitude parameter. 
The reverse sigmoid perturbation only replaces the final layer.
Then the final perturbed probability value is calculated as follows.
\begin{equation}
  \hat{y}_{j}^{i} = \alpha^{i}( y_{j}^{i} - \beta (s(\gamma s^{-1}(y_{j}^{i}))-1/2))
\end{equation}
where $\alpha^{i}$ is a sum-to-1 normalizer for $y^i$.

\subsection{Security Requirement against Model Extraction Attacks}

The goals of model extraction attacks can be divided into two objectives, i.e., accuracy, which measures the extracted model's correctness on test samples, and fidelity, which measures the agreement between the extracted model and original model on any point.
We measure the goal of the client's attack by accuracy because it is natural that the clients steal the model for use.
Informally, the goal of the defense against the client's model extraction attacks is to reduce the accuracy of the stolen model while maintaining the utility of the server's defense model. 
Formally, there are two metrics \cite{OrekondySF20} to measure the effectiveness of the defense.
\begin{itemize}
\item \textbf{\textit{Non-replicability.}}
The \textit{non-replicability} is measured in terms of the accuracy of the client's stolen model.
That is, the accuracy of the client's stolen model should be far lower than that of the server's model.

\item \textbf{\textit{Utility.}} The \textit{utility} is measured in terms of the accuracy of the server's model.
The defense method should have little impact on the server's model to maintain its utility.
\end{itemize}

To mitigate the model extraction attacks, the server locally trains a defense model and adds perturbations before the secure inference service starts.
This method is well compatible with the privacy-preserving inference protocol $\Pi_{\sys}$ and can be locally implemented by the server.
The server's model with defense is trained using the specific strategy described in~\S\ref{defensemethod}.
The server uses the defense model to provide inference service while mitigating the client’s model extraction attacks.

% %------------------------------------------------------------------------------

\subsection{{Evaluation}}\label{implementation}

\begin{figure*}[!ht]
  \centering
  \includegraphics[width=0.75\linewidth]{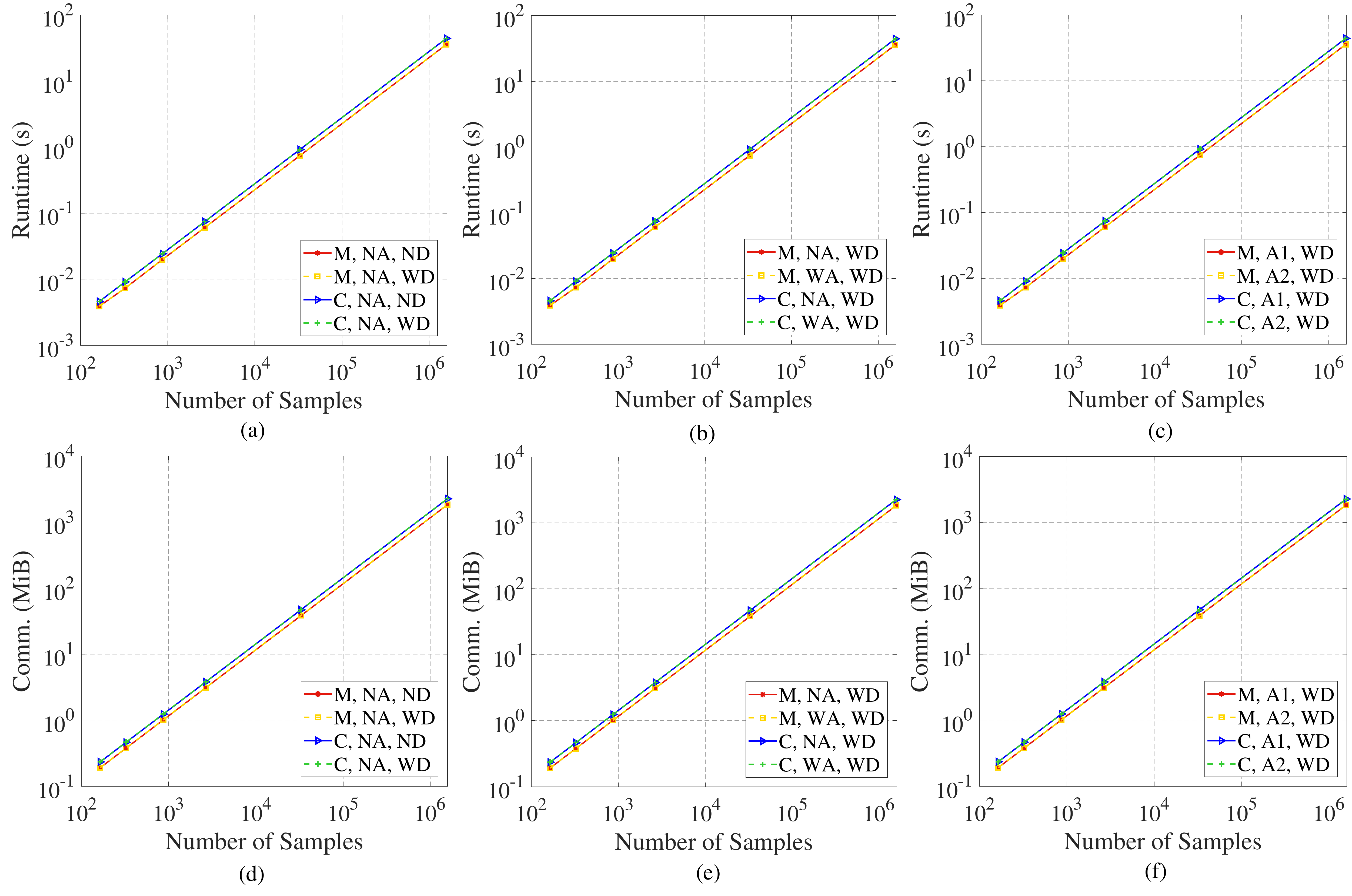}\\
  \caption{The impacts of different settings on the performance is depicted in this figure. M and C represent \textit{MNIST} and \textit{CIFAR-10} datasets, respectively. NA denotes no attack while WA denotes with attack. Similarly, ND and WD represent no defense and with defense respectively. A1 and A2 represent attack 1 and attack 2, respectively.}
  \label{attackimpact}
\end{figure*}

\noindent{\textbf{Experiment Setup:}}
Since the \sys integrated with a defense method can cause extra costs, we conducted experiments to test how much costs the defense will bring compared with those without defense.
Specifically, we modified the DNN network (in \S\ref{subsec:experimentalsetup}) by integrating it with the defense method (\S\ref{defensemethod}) and used it to train models (with defense ability) on both two datasets (\textit{MNIST} and \textit{CIFAR-10} datasets) respectively.
Similarly, we also used the same network without being changed (namely no defense ability) to train models on the two datasets.
Then we ran experiments by using these trained models and Cheetah as the underlying building block under different settings.
To compare the performance of \sys thoroughly, we ran experiments with extensive settings, i.e., the presence of the server’ defense, the presence of the client’s attack, as well as different attack methods (attack 1~\cite{PapernotMGJCS17} and attack 2~\cite{CarliniJM20}).
We change $R$ as $2^{3}$, $2^{5}$, $2^{7}$, $2^{9}$, $2^{13}$, and $2^{19}$, and the corresponding $B$ for each $R$ ranges from 8 to 3 (with fixed $T=100$). \looseness=-1

\noindent\textbf{(I) The Integrated Defense Has Negligible Impacts on \sys.} 
\autoref{attackimpact} shows the communication and runtime under different settings.
First, we compare the performances of \sys with defense (e.g., using the \textit{defense models}, denoted as WD) with that without defense (e.g., using the trained models with no defense, denoted as ND) to measure the \textbf{\textit{impact of the existence of defense.}}
As \autoref{attackimpact}. (a) and (d) show, the existence of defense brings very small extra costs when compared to without defense.
For example, when the total number of samples is 1,572,964, the \sys with defense increases only 3.956/8.488 MiB communication and 0.0007/0.0006 s runtime overhead compared to that without defense using dataset \textit{MNIST} and \textit{CIFAR-10} respectively, which is negligible. 
The reason for this is that the defense method only changes the final activation layer and adds very little computation compared to the overall computation.
Second, we compare the performances of \sys when there exist attacks (denoted as WA) with when there is no attack (NA) to measure the \textbf{{\textit{impact of the existence of attacks.}}}
As \autoref{attackimpact}. (b) and (e) show, the existence of the client’s attacks does not affect the running time and communication. 
Similarly, when the client uses different attack approaches (e.g., attack 1 (A1)~\cite{PapernotMGJCS17}, and attack 2 (A2)~\cite{CarliniJM20}), as shown in \autoref{attackimpact}. (b) and (e), the running time and communication remain nearly the same. 
The reason for this is that the client only exploits the attack after receiving the inference results, so it has no effect on the privacy-preserving inference computations.

\begin{figure*}[!ht]
  \centering
  \includegraphics[width=0.73\linewidth]{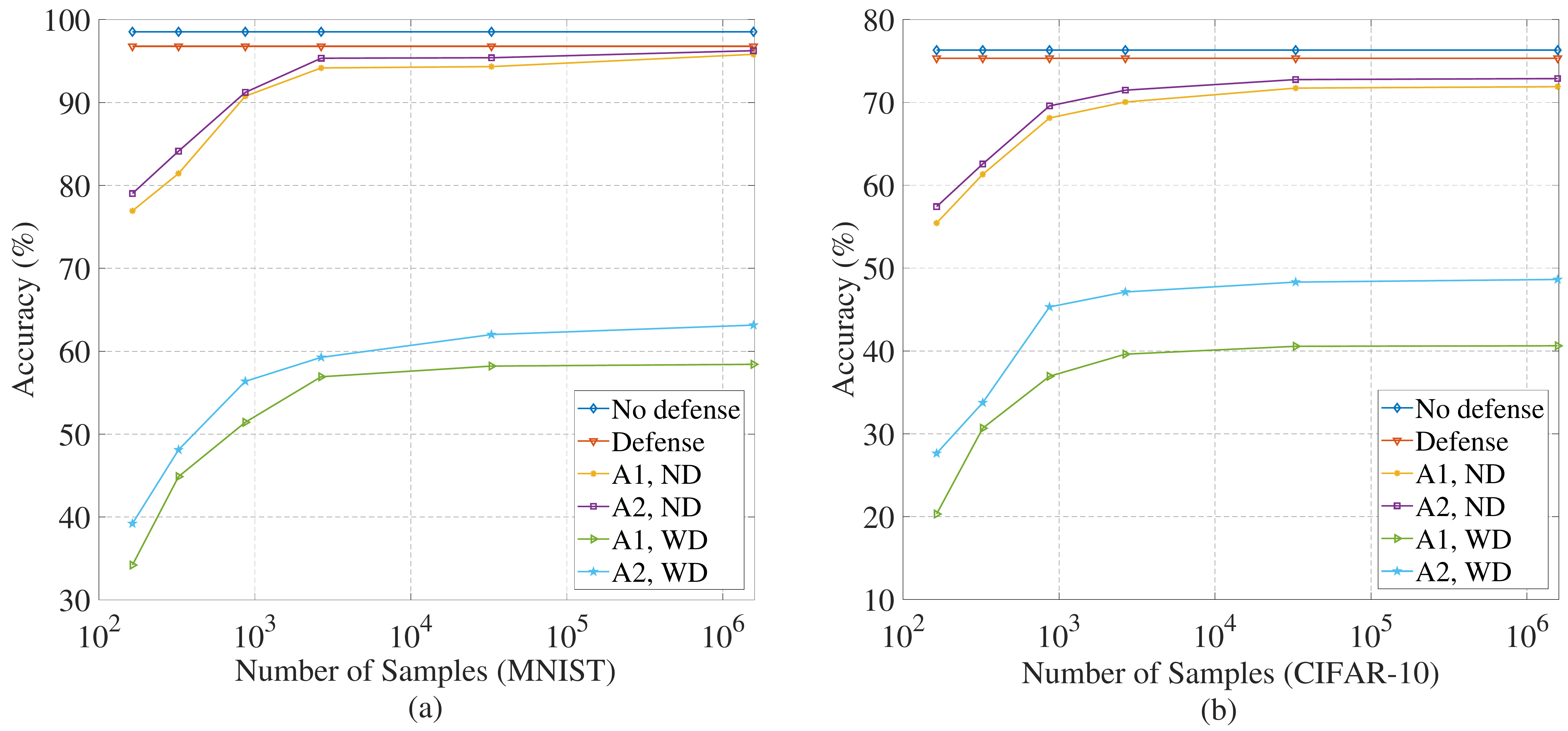}\\
  \caption{Accuracy of the client’s stolen models when the server uses models with defense and with no defense, the client uses different attack methods (A1 and A2), and the numbers of samples vary.}
  \label{defenseability}
  % \vspace{-5mm}
\end{figure*}

\vspace{1mm}
\noindent\textbf{(II) \sys Is Defense-Effective against Model Extraction Attacks.}
We would like to test the defense effectiveness of \sys against model extraction attacks.
We conducted experiments by letting the server use the models with defense (WD) and no defense (ND), and the client perform two attack approaches (i.e., attack 1 (A1)~\cite{PapernotMGJCS17} and attack 2 (A2)~\cite{CarliniJM20}) respectively when the total numbers of query samples vary.
To test the {\emph{non-replicability}} of \sys, we tested the model accuracy of the client's stolen models.

As shown in \autoref{defenseability}, the accuracy of the client's stolen models significantly decreases when the server used defense compared with when there is no defense, indicating the  {\emph{non-replicability}} of \sys.
In other words, \sys's defense is effective on both two datasets and when the client performed two different attacks.
When using the \textit{MNIST} dataset and the numbers of total samples vary, if the server used the models with defense, the accuracy of the client’s stolen models decreases by 36.1-42.7\% (A1), while the accuracy decreases by 33.1-39.8\% (A2).
For instance, when the number of samples is 1,572,964, the accuracy of the client’s stolen model drops from 95.81\% (A1, ND), 96.25\% (A2, ND) to 58.42\% (A1, WD), 63.15\% (A2, WD), respectively.
When using the \textit{CIFAR-10} dataset and the total number of samples is 1,572,964, for example, the accuracy of the client’s stolen model drops from 71.9\% (A1, ND), 72.88\% (A2, ND) to 40.62\% (A1, WD), 48.63\% (A2, WD), respectively.\vspace{-0.3mm}

\vspace{1mm}
\noindent\textbf{(III) \sys Maintains \textit{Utility} When Integrated with the Defense.} 
To evaluate the \textit{utility} of the server's model with defense, we tested the model accuracy of the models with defense and with no defense on two datasets respectively.
As \autoref{defenseability} depicts, \sys maintains the \textit{utility}, e.g., the accuracy of the server’s models with defense (labelled as Defense in \autoref{defenseability}) slightly decreases when compared with those with no defense (No defense).
For instance, when using the \textit{MNIST} and \textit{CIFAR-10} datasets, the model accuracy of the models with defense is 96.78\% and 75.34\%, while that with no defense is 98.53\% and 76.31\%, respectively.\looseness=-1
\end{appendices}

\end{document}